\documentclass[a4paper,fleqn,usenatbib]{mnras}

\usepackage[T1]{fontenc}
\usepackage{ae,aecompl}

\usepackage{graphicx}	
\usepackage{amsmath}	
\usepackage{amssymb}	
\usepackage{newtxtext,newtxmath}
\newcommand{\nicer}{\textit{NICER}~}
\newcommand{\nustar}{\textit{NuSTAR}~}

\title[MAXI J0637--430]{{\it NICER} observations of the black hole candidate MAXI~J0637--430 during the 2019-2020 Outburst}

\author[Jana et al.]{
 Arghajit Jana$^{1}$\thanks{E-mail: argha@prl.res.in},
 Gaurava K. Jaisawal$^{2}$,
 Sachindra Naik$^{1}$,
 Neeraj Kumari$^{1,3}$, 
 \newauthor
 Birendra Chhotaray$^{1,3}$,
 Diego Altamirano$^{4}$,
 Ronald A. Remillard$^5$, 
 Keith C. Gendreau$^{6}$ 
\\
 $^{1}$Astronomy \& Astrophysics Division, Physical Research Laboratory, Navrangpura, Ahmedabad, 380009, India\\
 $^{2}$National Space Institute, Technical University of Denmark, Elektrovej, 327-328, DK-2800 Lyngby, Denmark \\
 $^3$Indian Institute of Technology, Gandhinagar - 382355, Gujarat, India\\
 $^{4}$School of Physics and Astronomy, University of Southampton, Southampton, SO17 1BJ, UK\\ 
 $^5$MIT Kavli Institute for Astrophysics and Space Research, MIT, 70 Vassar Street, Cambridge, MA 02139, USA\\
 $^6$Astrophysics Science Division, NASA Goddard Space Flight Center, Greenbelt, MD 20771, USA\\
 }
\date{Accepted XXX. Received YYY; in original form ZZZ}

\pubyear{2020}

\begin{document}

\label{firstpage}
\pagerange{\pageref{firstpage}--\pageref{lastpage}}
\maketitle

\begin{abstract}
We present detailed timing and spectral studies of the black hole candidate MAXI~J0637--430 during its 2019-2020 outburst using observations with the {\em Neutron Star Interior Composition Explorer (NICER)} and the {\em Neil Gehrels  Swift Observatory}. We find that the source evolves through the soft-intermediate, high-soft, hard-intermediate and low-hard states during the outburst. No evidence of quasi-periodic oscillations is found in the power density spectra of the source. Weak variability with fractional rms amplitude $<5\%$ is found in the softer spectral states. In the hard-intermediate and hard states, high variability with the fractional rms amplitude of $>20\%$ is observed. The $0.7-10$~keV spectra with {\em NICER} are studied with a combined disk-blackbody and nthcomp model along with the interstellar absorption. The temperature of the disc is estimated to be $0.6$~keV in the rising phase and decreased slowly to $0.1$~keV in the declining phase. The disc component was not detectable or absent during the low hard state. From the state-transition luminosity and the inner edge of the accretion flow, we estimate the mass of the black hole to be in the range of $5-12$ $M_{\sun}$, assuming the source distance of $d<10$~kpc.
\end{abstract}



\section{Introduction}
\label{sec:intro}

An X-ray binary (XRB) consists of a normal star and a compact object. The compact object can be a black hole (BH), or a neutron star (NS). Depending on the mass of the companion star, the XRBs can be classified as a high mass X-ray binary (HMXB) or a low mass X-ray binary (LMXB) \citep{White1995,RM06}. An HMXB system contains an O- or B-type companion star, while an LMXB contains an A-type or later star \citep{Tetarenko2016}. A transient XRB spends most of the time in the quiescent state during which the compact object is marginally detectable or even non-detectable with the current generation X-ray detectors. The transient XRBs occasionally show X-ray outbursts that last for several weeks to months. During the outburst, the X-ray luminosity of the source increases by several orders of magnitude compared to the quiescent state.

A spectrum of a black hole X-ray binary (BHXRB) can be approximated with a soft thermal multi-colour blackbody component and a non-thermal power-law component. The multi-colour blackbody component originates from a standard thin accretion disc \citep{SS73,NT73}. In contrast, the power-law component originates in a Compton cloud located close to the BH \citep{ST80,ST85}. The soft X-ray photons originated from the standard accretion disc undergo inverse-Comptonization in the Compton cloud and produce the hard power-law component \citep{HM93,Z93,T94,CT95,Zycki1999,Done2007}.

An outbursting BHXRB shows rapid variation and fluctuation in spectral and timing properties  \citep{Mendez1997,vanderklis1989,vanderklis2000}. A correlation between the spectral and timing properties of the source can be seen in the hardness--intensity diagram \citep[HID;][]{Homan2001,Homan2005,Nandi2012}, accretion rate--intensity diagram \citep[ARRID;][]{AJ2016}, rms--intensity diagram  \citep[RID;][]{Munoz-Darias2011}, or hardness--rms diagram \citep[HRD;][]{Belloni2005}. In general, an outbursting BHXRB exhibits four different spectral states, viz. low hard state (LHS), hard-intermediate state (HIMS), soft-intermediate state (SIMS) and high soft state (HSS), and evolves as LHS --> HIMS --> SIMS --> HSS --> SIMS --> HIMS --> LHS \citep{RM06,Nandi2012,AJ2020b}. A BHXRB also shows low-frequency quasi-periodic oscillations (LFQPOs) in the power density spectra (PDS) observed in a range of 0.1--20 Hz. A LFQPO can be classified as type-A, type-B or type-C depending on the Q-value of the QPO (Q=$\nu/\Delta \nu$, $\nu$ and $\Delta \nu$ are centroid QPO frequency and full-width-half-maxima, respectively), nature of broadband noise, and rms amplitude of the QPO \& broad noise \citep[][ and references therein]{Casella2005}. 

Each spectral state is characterized by different spectral and timing properties \citep[for a review, see;][]{RM06}. The LHS is characterized by a cool disc of temperature in $\sim 0.2-0.5$~keV range and photon index, $\Gamma \sim 1.5-1.7$. Sometimes, the disc component is not detectable in this state. In general, the hard X-ray photon flux dominates over the soft X-ray photon flux. Evolving type-C QPO is observed in this state. Along with the QPO, broadband noise and rms amplitude of $\sim 20-40$ \% are observed in the PDS of the source. A compact and quasi-stable jet is also observed in the LHS \citep{Fender2004}. In the HIMS, the soft photon flux increases relative to the hard photon flux. The source spectra became steep with a photon index, $\Gamma \sim 2$. Evolving type-C QPO is also observed in this state. The SIMS is associated with $\Gamma \sim 2.2-2.5$, with high soft photon flux. Sporadic type-A or type-B QPOs are observed in this state. The SIMS is often associated with a discrete ejection or blobby jets. The HSS is dominated by the disc flux or soft X-ray photon flux, with temperature $T \sim 1$~keV. The spectra are observed to be steep with $\Gamma \geq 2.5$. No QPO is observed in this state. A week broadband noise with rms amplitude $< 5$\% is observed in this state. No jet is observed in the HSS \citep{Fender2004}.

The black hole candidate (BHC) MAXI~J0637--430 was discovered with {\it MAXI}/GSC on 2 November 2019 during the onset of the 2019-2020 X-ray outburst \citep{Negoro19}. The outburst continued for $\sim 6$ months. The {\it Swift}/XRT observation of the field localized the source at RA/Dec(J2000) = 99.09828\textdegree, -42.8678\textdegree \citep{Kennea19}. After the discovery, the source was observed in optical \citep{Li19}, infrared \citep{Murata19}, and radio wavelengths \citep{Russell19}. Several X-ray satellites such as {\it NICER} \citep{Remillard20}, {\it AstroSAT} \citep{Thomas19}, and {\it NuSTAR} \citep{Tomsick19} also reported the primary timing and spectral analysis of the source. Preliminary studies suggest that the source is an LMXB \citep{Strader19} hosting a BH as the compact object \citep{Tomsick19}.

In this paper, we present our studies on the 2019-2020 X-ray outburst of MAXI~J0637--430 using data from the {\it NICER} and {\it Swift} observatories. The paper is organized in the following way. In \S2, we describe the observation and data analysis processes. The results obtained from our timing and spectral analysis are presented in \S3. In \S4, we discuss our findings and finally, in \S5, we summarize our results.

\begin{figure*}
\vskip 0.2cm
\centering
\includegraphics[width=16cm,keepaspectratio=true]{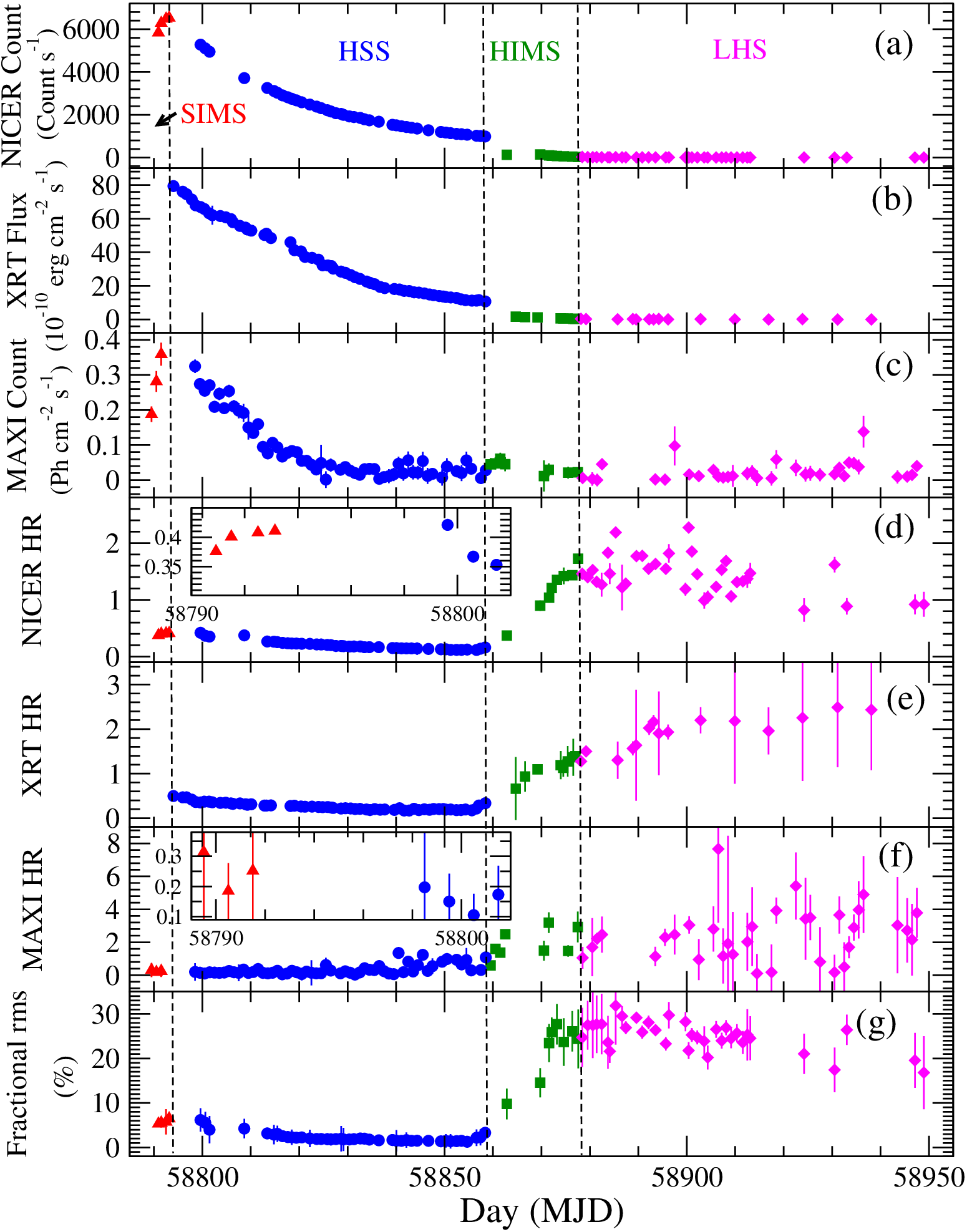}
\caption{The variation of (a) $0.5-10$~keV {\it NICER} count rate in counts~s$^{-1}$, (b) $0.5-10$~keV {\it Swift}/XRT flux in $10^{-10}$ erg cm$^{-2}$ s$^{-1}$, (c) $2-20$~keV {\it MAXI}/GSC count rate in photons cm$^{-2}$ s$^{-1}$, (d) {\it NICER} hardness ratio (HR-1), (e) {\it Swift}/XRT hardness ratio (HR-2), and (f) {\it MAXI}/GSC hardness ratio (HR-3) are shown with days (in MJD). In panel (g), the evolution of $0.1-50$~Hz fractional rms amplitude in the $0.5-10$~keV energy band, obtained from {\it NICER} observation, is shown. The red triangles, blue circles, green squares, and magenta diamond indicate SIMS, HSS, HIMS, and LHS, respectively, during the outburst. The vertical dashed lines separates different spectral states.}
\label{fig:lc}
\end{figure*}

\begin{figure*}
\includegraphics[width=7.5cm]{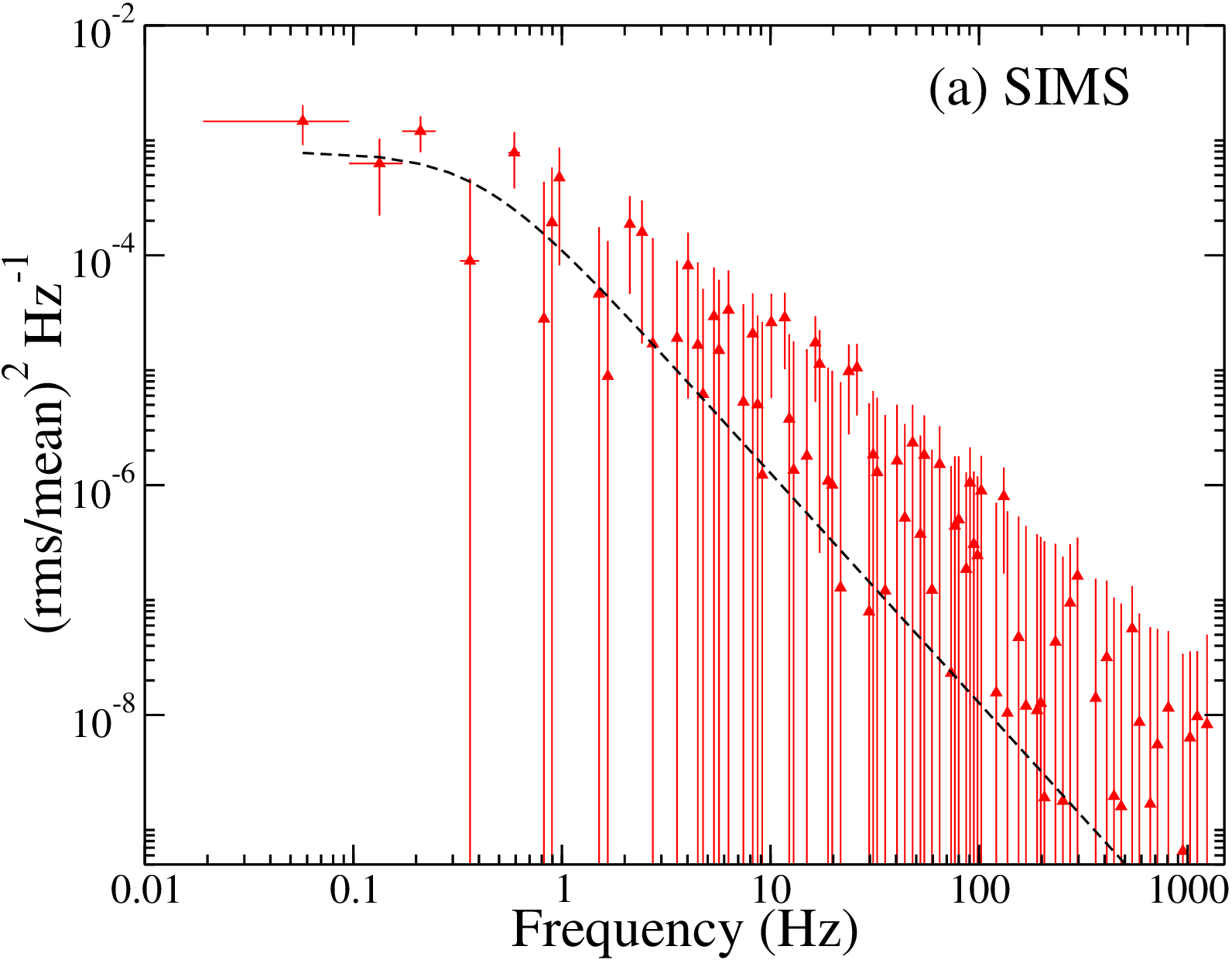} \hskip0.2cm
\includegraphics[width=7.5cm]{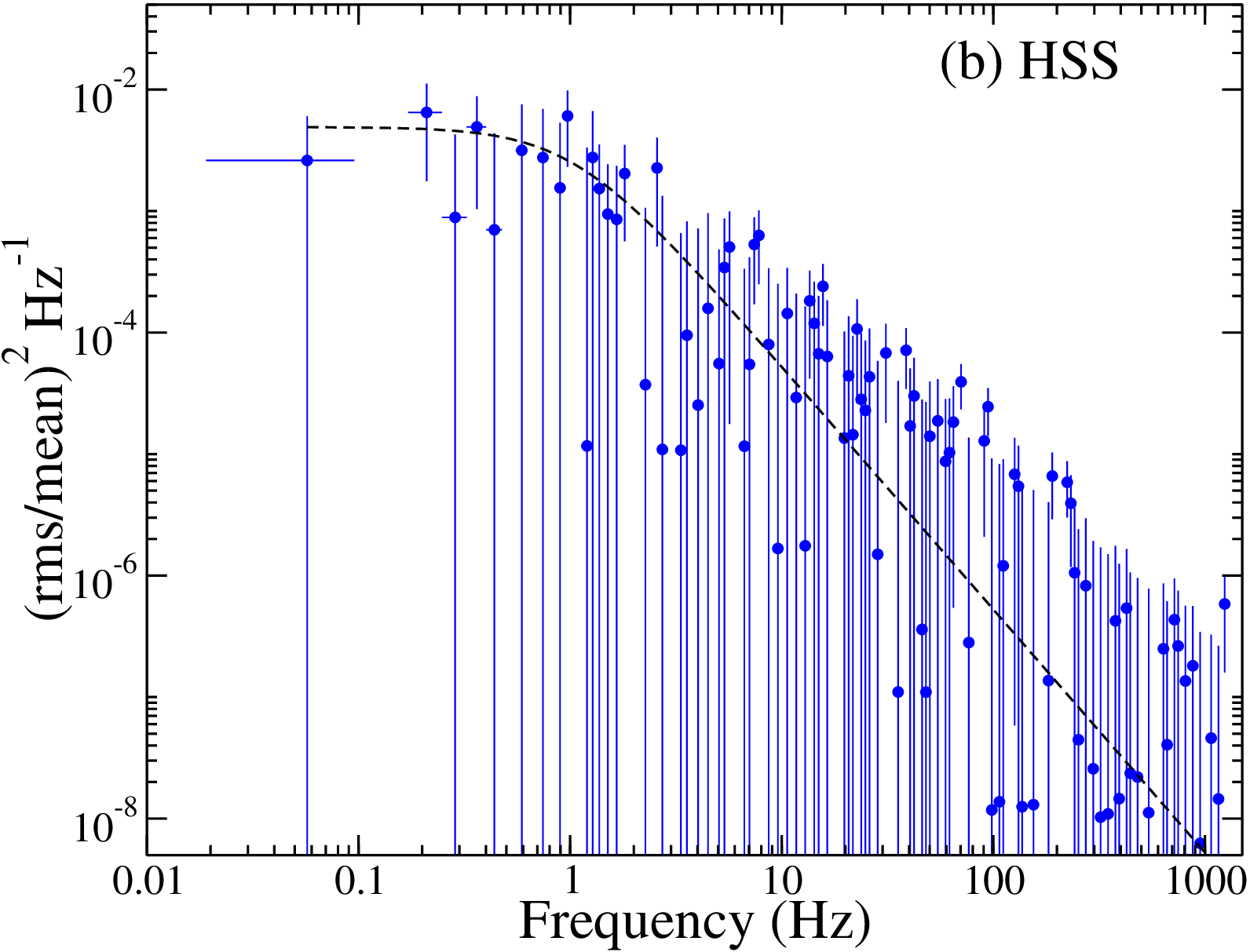} \\
\vskip0.2cm
\includegraphics[width=7.5cm]{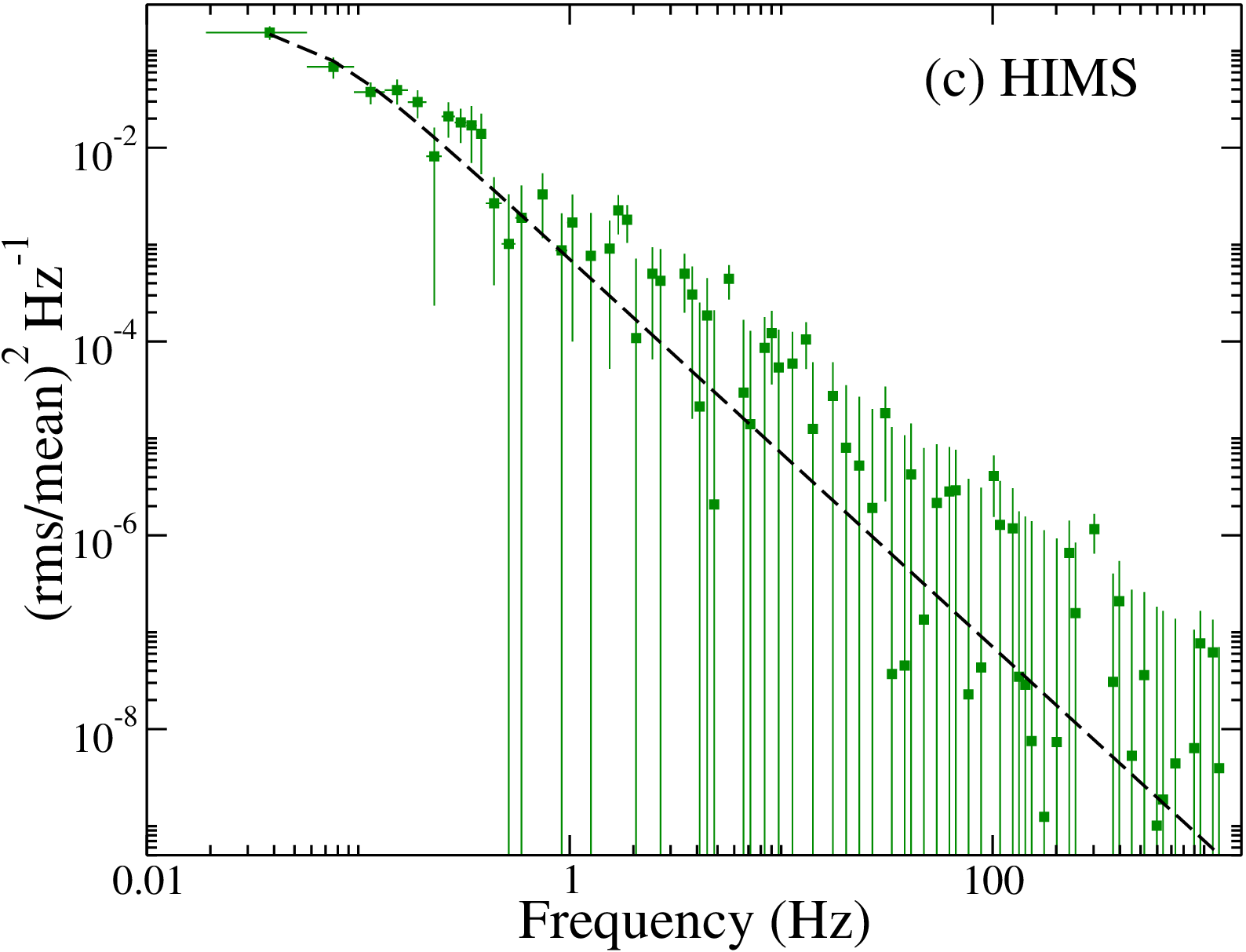} \hskip0.2cm
\includegraphics[width=7.5cm]{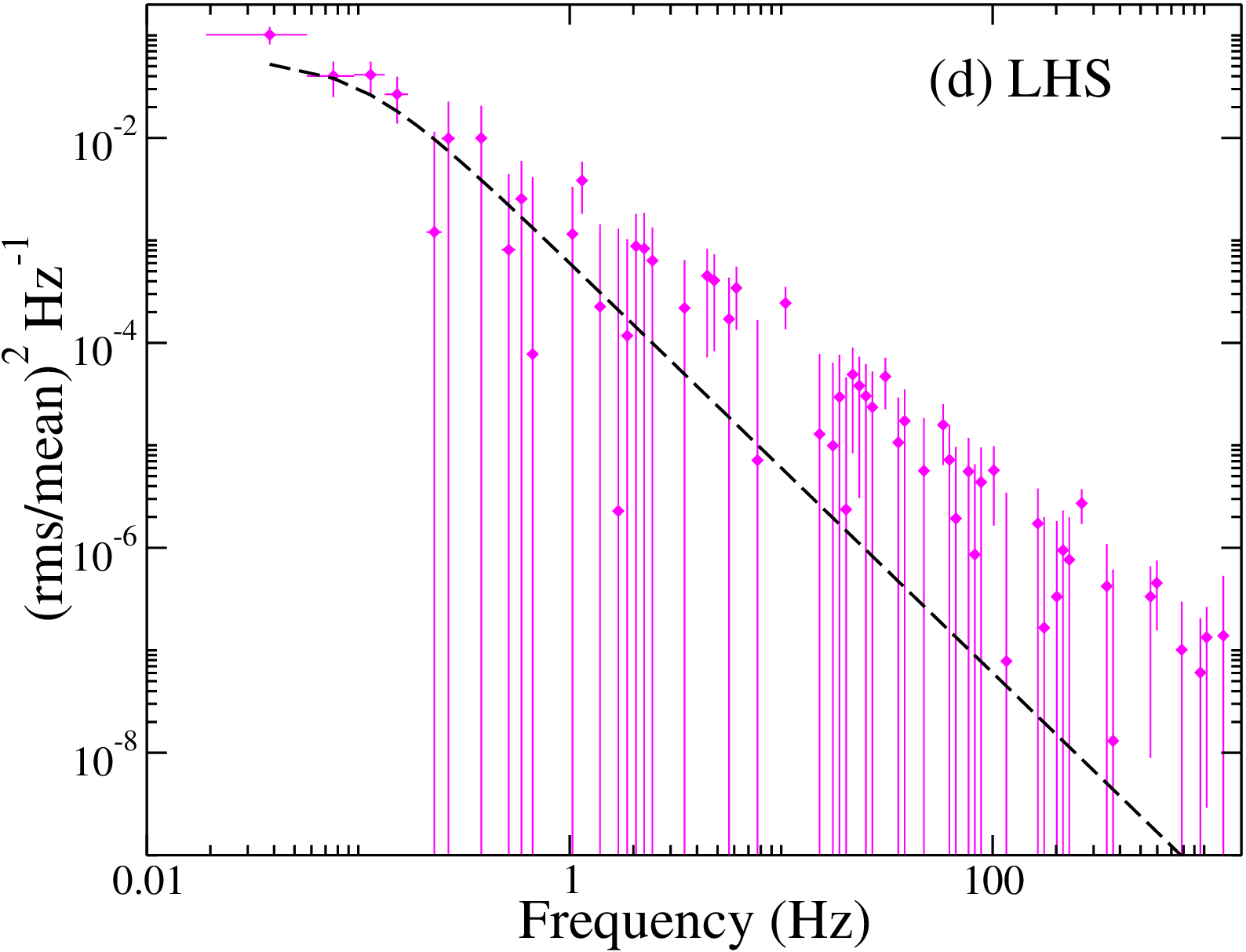}
\caption{Power density spectra obtained from the \nicer observations on (a) 2019 November 4 (Obs ID : 2200950102), (b) 2020 January 2 (Obs ID: 2200950145), (c) 2020 February 10 (Obs ID: 2200950173), and (d) 2020 March 2 (Obs ID: 3200950102). The PDS are fitted with Lorentizian function. The dashed lines represent the best-fit Lorentzian.}
\label{fig:pds}
\end{figure*}

\begin{figure}
\includegraphics[width=8.5cm]{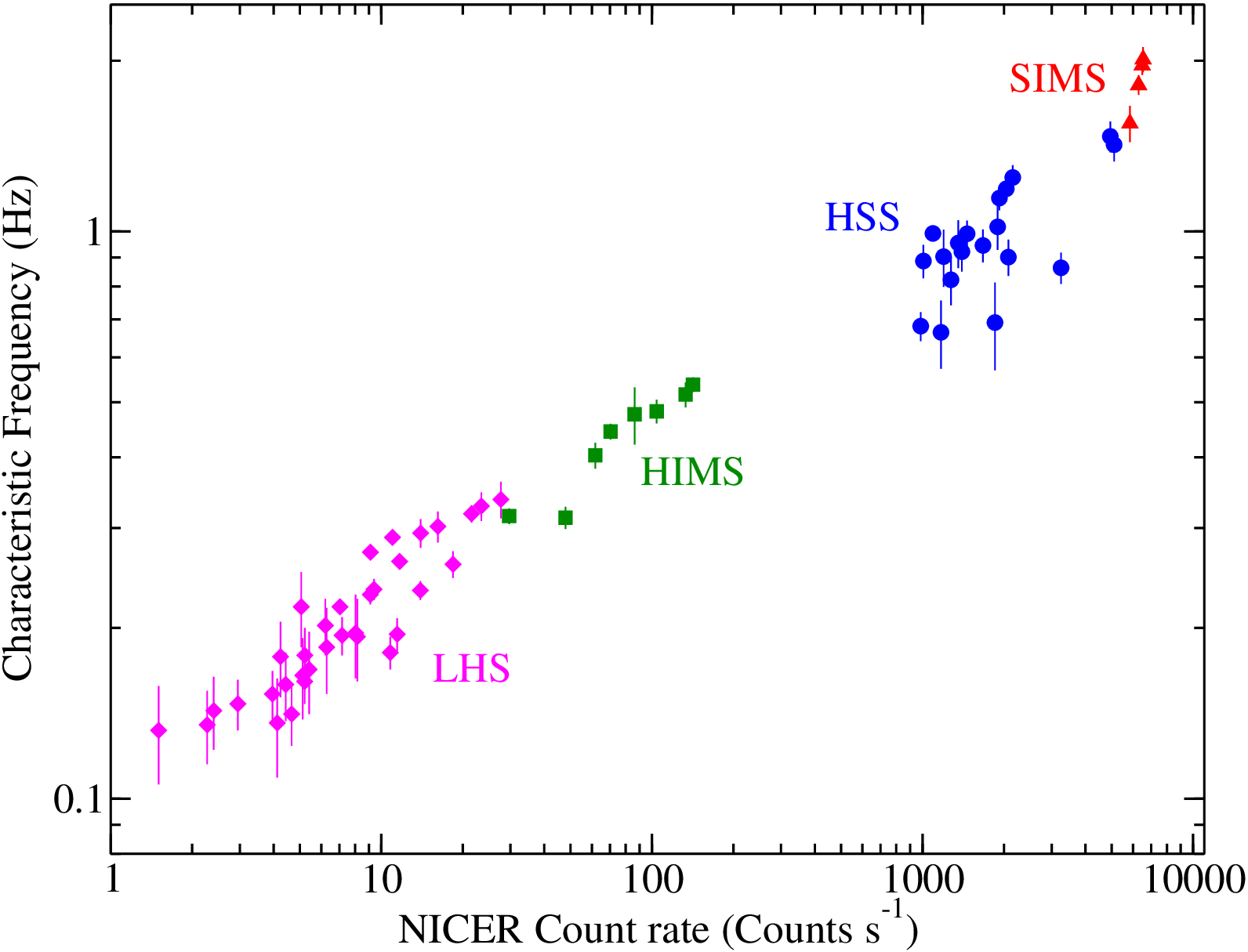}
\caption{The characteristic frequency ($\nu_c$) of broadband noise component is plotted as a function of $0.5-10$~keV {\it NICER} count rate. The colours and symbol correspond to the different spectral states of the source during the outburst as described in Fig.~\ref{fig:lc}.}
\label{fig:charac-freq}
\end{figure}

\section{Observation and Data Analysis}
\label{sec:obs}

\subsection{\it NICER}
\label{sec:nicer}

Soon after its discovery, the BHC MAXI~J0637--430 was observed with {\it NICER} at several epochs between 3 November 2019 and 9 April 2020. {\it NICER} is an attached external payload on the International Space Station that offers an X-ray timing instrument  \citep[XTI;][]{Gendreau2012} working in the 0.2--12~keV photon energy range. The XTI consists of 56 X-ray ``concentrator'' optics, each associated with a silicon drift detector \citep{Prigozhin2012}. There are 52 active detectors, providing a total effective area of 1900 cm$^2$ at 1.5 keV. The timing resolution of \nicer is $\sim 100$ ns, whereas the spectral resolution is $\sim 85$~eV at $1$~keV. We used a total of 88 epochs of observations under the IDs 22009501xx and 32009501xx (see~Table~\ref{tab:analysis}) with a total exposure time of $\approx 137$~ks to study the nature of the BHC MAXI~J0637--430 during its 2019--2020 X-ray outburst. The data were reprocessed with the {\tt nicerl2} \footnote{\url{https://heasarc.gsfc.nasa.gov/docs/nicer/analysis_threads/nicerl2/}} script in the presence of the latest gain and calibration files of version 20200722. Standard GTIs were also generated using the {\tt nimaketime} task. The reprocessed cleaned events were used for extracting the light curves and spectra in the {\tt XSELECT} environment of {\tt FTOOLS}. Ancillary response file and response matrix file of version 20200722 are considered in our spectral analysis. The background corresponding to each epoch of the observations is simulated by using the {\tt nibackgen3C50}\footnote{\url{https://heasarc.gsfc.nasa.gov/docs/nicer/toolsnicer_bkg_est_tools.html}} tool (Remillard et al., in prep.). 

The $0.5-10$~keV light curves were created at a 400 $\mu$s time bin. Power density spectra (PDS) were generated by applying the Fast Fourier Transformation (FFT) technique on the light curves using {\tt powspec} task of {\tt FTOOLS}. We divided the light curves into 8192 intervals and computed the Poison noise subtracted PDS for each interval. Then, we averaged all the PDS to obtain the final PDS for each observation. The final PDS are normalized to give the fractional rms spectra in $(rms/mean)^2 Hz^{-1}$ unit. Then, we re-binned the PDS with a factor of 1.05.

For the spectral study, we used data in the energy range of $0.7-10$~keV. The spectra were re-binned to have a minimum of 20 counts per bin. Data below 0.7 keV are not included in our spectral fitting due to observed calibration issues like excess at lower energies. We added a systematic of 1.5\% during the spectral analysis, as recommended by the \nicer team\footnote{\url{https://heasarc.gsfc.nasa.gov/docs/nicer/data_analysis/nicer_analysis_tips.html}} \citep{Jaisawal2019}.

\subsection{\it Swift}
\label{sec:swift}

The {\it Swift}/XRT observed MAXI~J0637--430 several times between 7 November 2019 and 10 June 2020. We used a total of 85 epochs of observations, observed between 7 November 2019 and 30 March 2020. The {\it Swift}/XRT observations were carried out in both window-timing (WT) and photon counting (PC) mode, under the IDs 0001217, and 00088999 with a total exposure time of $\sim 100$~ks. The $0.5-10$~keV spectra were generated using the standard online tools provided by the UK {\it Swift} Science Data Centre \citep{Evans2009}\footnote{\url{http://www.swift.ac.uk/user_objects/}}. For the present study, we used both WT and PC mode spectra and light curves in the energy band of $0.5-10$~keV.

\section{Result}
\label{sec:res}

\subsection{Outburst Profile}
\label{sec:profile}

Following the discovery, MAXI~J0637--430 was extensively observed with {\it NICER} and {\it Swift}/XRT. In Fig.~\ref{fig:lc}, we show the evolution of the $0.5-10$~keV {\it NICER} light curve in counts~s$^{-1}$ (panel-a), the $0.5-10$~keV {\it Swift}/XRT flux in $10^{-10}$ erg cm$^{-2}$ s$^{-1}$ (panel-b), and the $2-20$~keV {\it MAXI}/GSC count rate in photons cm$^{-2}$ s$^{-1}$ (panel-c). The hardness ratios obtained from the {\it NICER} observations (panel-d), {\it Swift}/XRT observations (panel-e) and {\it MAXI}/GSC observations (panel-f) are shown in Fig~\ref{fig:lc}. The $0.01-50$~Hz fractional rms amplitude ($r)$ in $0.5-10$~keV energy band, obtained from the {\it NICER} observations during the outburst, is also shown in the bottom panel (panel-g) of Fig~\ref{fig:lc}. The insets in panels-d \& f represent the magnified version of the hardness ratios from the {\it NICER} and {\it MAXI}/GSC observations during the peak of the outburst. The hardness ratio (HR) for the {\it NICER} observations is obtained by taking ratio between the count rates in $2-10$~keV and $0.5-2$~keV energy bands. For the {\it Swift}/XRT observations, the HR is defined as the ratio between the $2-10$~keV flux and $0.5-2$~keV flux. For {\it MAXI}/GSC observations, the HR is defined as the ratio between the count rates in $4-20$~keV and $2-4$~keV energy bands. 

As mentioned earlier, {\it NICER} started observing the source one day after the discovery, i.e. from 3 November 2019 (MJD 58790.92). The X-ray intensity of the source rapidly increased and became maximum on 6 November 2019 (MJD 58793.14) with a count rate of 6533 count s$^{-1}$ (Fig.~\ref{fig:lc}(a)). After reaching the peak on 6 November 2019, the source entered the declining phase of the outburst. {\it Swift}/XRT started observing the source from 7 November 2019 November (MJD 58794.03), when the source was already in the declining phase (Fig.~\ref{fig:lc}(b)). From this day, the X-ray intensity gradually decreased up to 10 January 2020 (MJD 58858.37). After this, the X-ray intensity suddenly decreased by $\sim$10 times on 14 January 2020 (MJD  58862.85) followed by a gradual decrease (Fig.~\ref{fig:lc}(a) \& (b)). After that, the source faded toward the quiescent state. The $0.5-10$~keV {\it Swift}/XRT flux and $2-20$~keV {\it MAXI}/GSC count rate also showed similar variation. From the top three panels of Fig.~\ref{fig:lc}, the outburst can be classified as the fast-rise-exponential decay type \citep{Chen1997}.

We observed that the {\it NICER} HR increased gradually in the rising phase of the outburst (inset figure in Fig.~\ref{fig:lc}(d) and reached the maximum on 12 November 2019 with HR $\sim 0.4$. After that, the HR decreased slowly and became steady at a value of $\sim 0.12$, till 10 January 2020 (MJD 58858.37). From 14 January 2020, the HR increased again until 29 January 2020 (MJD 58877.58) and remained high (in the range of 1 and 2) till the end of the observations. The HR obtained from the {\it Swift}/XRT and {\it MAXI}/GSC observations also showed similar behaviour. From the evolution of the HRs obtained from three instruments and fluxes in the soft and hard X-ray bands, we classified the entire outburst in four spectral states as marked in different colours and symbols in Fig~\ref{fig:lc} : soft-intermediate state (SIMS - marked in red triangles) in the rising phase, high soft state (HSS - marked in blue circles), hard-intermediate state (HIMS - marked in green squares) in the declining phase, and low hard state (LHS - marked as magenta diamonds).

\begin{figure}
\includegraphics[width=8.5cm,keepaspectratio=true]{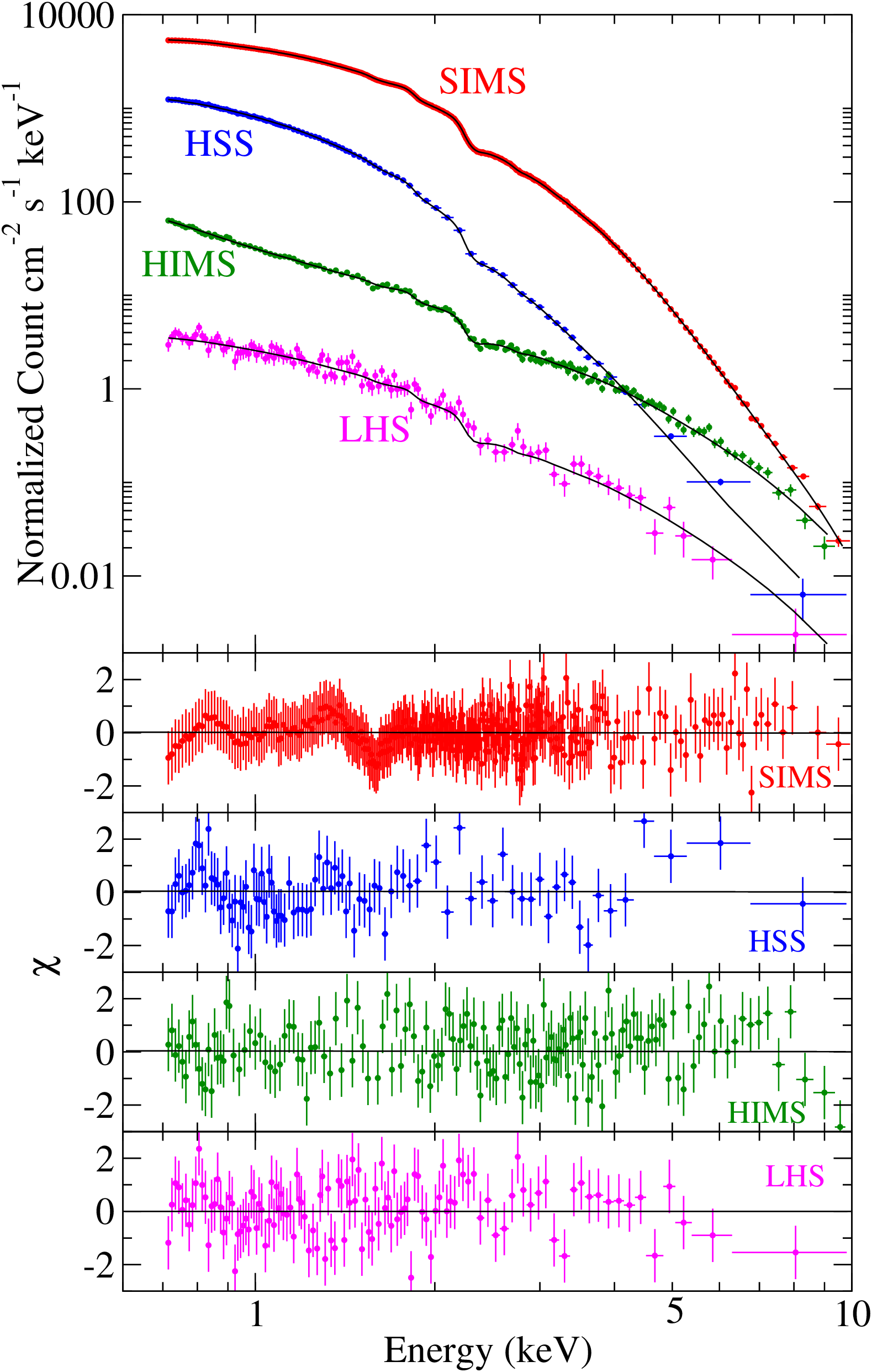}
\caption{Energy spectra of MAXI~J0637--430, obtained from the {\it NICER} observations at four different phases of the outburst, are shown in the top panel. Corresponding residuals obtained by fitting the spectra with disk blackbody and power law model, along with the interstellar absorption, are shown in the bottom panels. Red triangles, blue circles, green squares, and magenta diamonds correspond to the spectra observed on 4 November 2019 (Obs ID : 2200950102 -- SIMS in the rising phase), 2 January 2020 (Obs ID: 2200950145 -- HSS), 10 February 2020 (Obs ID: 2200950173 -- HIMS in the declining phase) and 2 March 2020 (Obs ID: 3200950102 -- LHS in the declining phase), respectively. In the upper panel, the black lines represent the best-fit model.}
\label{fig:spec}
\end{figure}

\subsection{Power Density Spectra}
\label{sec:pds}

The PDS are generated from all \nicer light curves to investigate the characteristics of the profiles. The light curves with 400 $\mu$s time resolution allowed us to search for the QPO features in the PDS up to a maximum frequency of $1250$~Hz. Representative PDS obtained from the {\it NICER} observations on 4 November 2019 (Obs ID : 2200950102, panel-a), 2 January 2020 (Obs ID: 2200950145, panel-b), 10 February 2020 (Obs ID: 2200950173, panel-c) and 2 March 2020 (Obs ID: 3200950102, panel-d) are shown in Fig.~\ref{fig:pds}. We calculated the fractional rms amplitude ($r$) of the PDS by integrating the power in $0.01-50$~Hz range \citep{vanderklis2004}. We fitted the PDS with multiple Lorentzian functions. Most of the PDS in the SIMS and HSS were found to be fitted with a single zero-centroid Lorentzian, while more than one Lorentzian functions were required to fit the PDS in the HIMS and LHS. We calculated the characteristic frequency ($\nu_{\rm C}$) of the broadband noise for those observations. The characteristic frequency is the frequency where the component contributes the most of its variance per frequency and given by,  $\nu_{\rm C} = \sqrt{\nu_{\rm 0} + (\Delta \nu/2)^2}$, where $\nu_{\rm 0}$ and $\Delta \nu$ are the centroid frequency and full-width-half maxima, respectively \citep{Nowak2000,Belloni2002}.

We searched for QPO in the PDS and found none. Thus, we tried to estimate the upper limit of the QPO. We found peaked noise in the PDS with an upper limit (UL) of rms $<5\%$ at $1\sigma$ confidence level in five observations. As those peaked noises were detected at $<1\sigma$ level, we discard these band-limited peaked noises as QPO. We also tried to find QPO by combining the PDS from several observations in a particular state. However, we did not detect any evidence of QPO. We plotted the characteristic frequency as a function of $0.5-10$~keV NICER count rate and found that they are correlated. We found weak red noise in the rising phase of the outburst ($r \sim 5 \%$). The strength of the signal monotonically decreased to $r \sim 1 \%$ in the declining phase until 10 January 2010 (MJD 58858.37). After that, the fractional rms amplitude increased to $r \sim 23 \%$ on 23 January 2020. Then, the fractional rms amplitude varied within $r \sim 20-30$ \% till the end of the outburst (see Fig.~\ref{fig:lc}g). In this phase, the PDS showed weak flat-top noise.

\begin{figure*}
\vskip 0.2cm
\centering
\includegraphics[width=16cm,keepaspectratio=true]{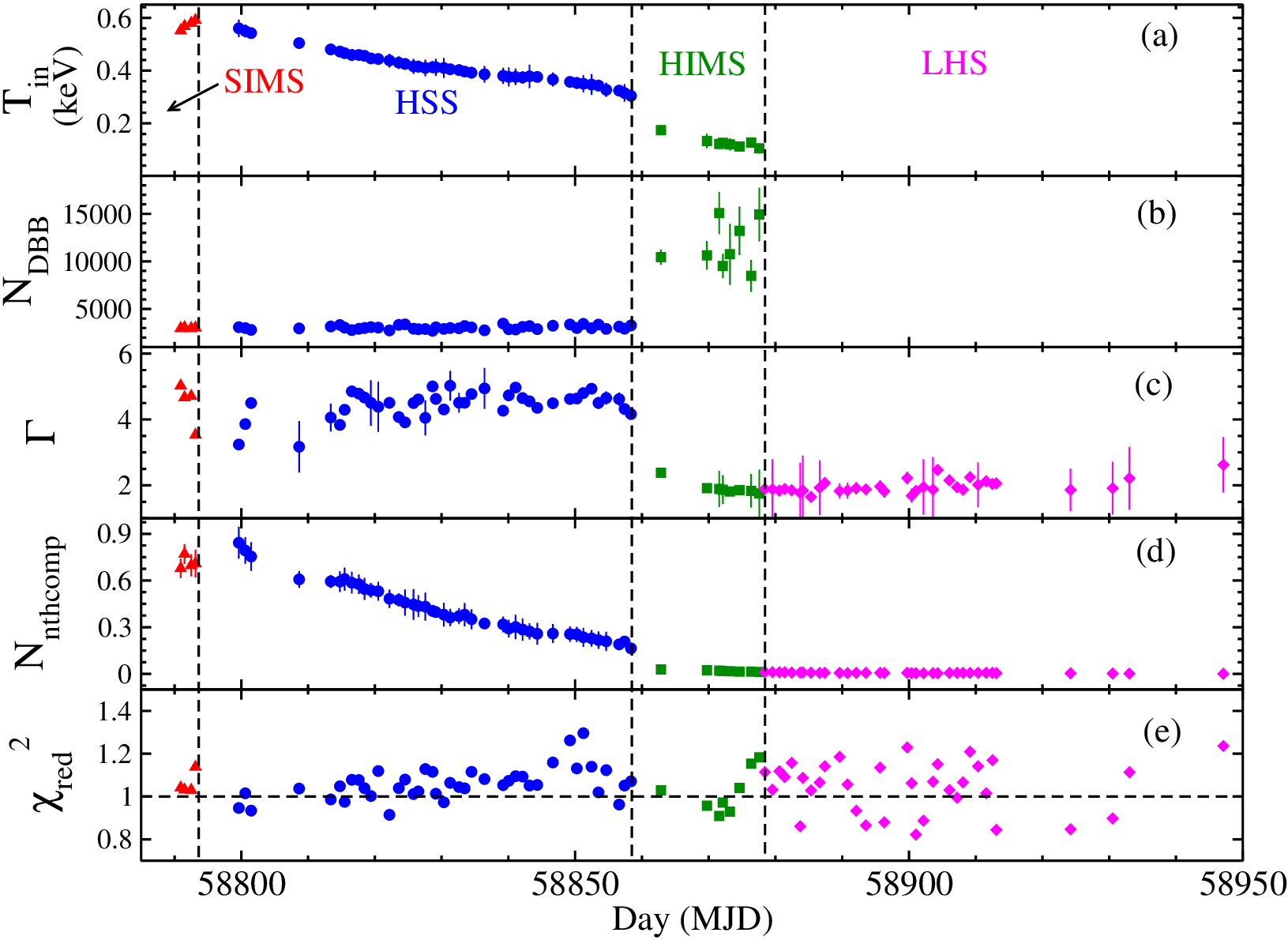}
\caption{The evolution of (a) the inner disc temperature ($T_{\rm in}$, in keV), (b) the diskbb normalization ($N_{\rm DBB}$), (c) the photon index ($\Gamma$), (d) the nthcomp normalization ($N_{\rm nthcomp}$), and (e) the reduced-$\chi^2$ ($\chi^2_{\rm red} = \chi^2$/dof) are shown. The colours and symbol correspond to the different spectral states of the source during the outburst as described in Fig.~\ref{fig:lc}. The vertical dashed lines separate different spectral states.}
\label{fig:spec-param}
\end{figure*}

\begin{figure*}
\vskip 0.2cm
\centering
\includegraphics[width=16cm,keepaspectratio=true]{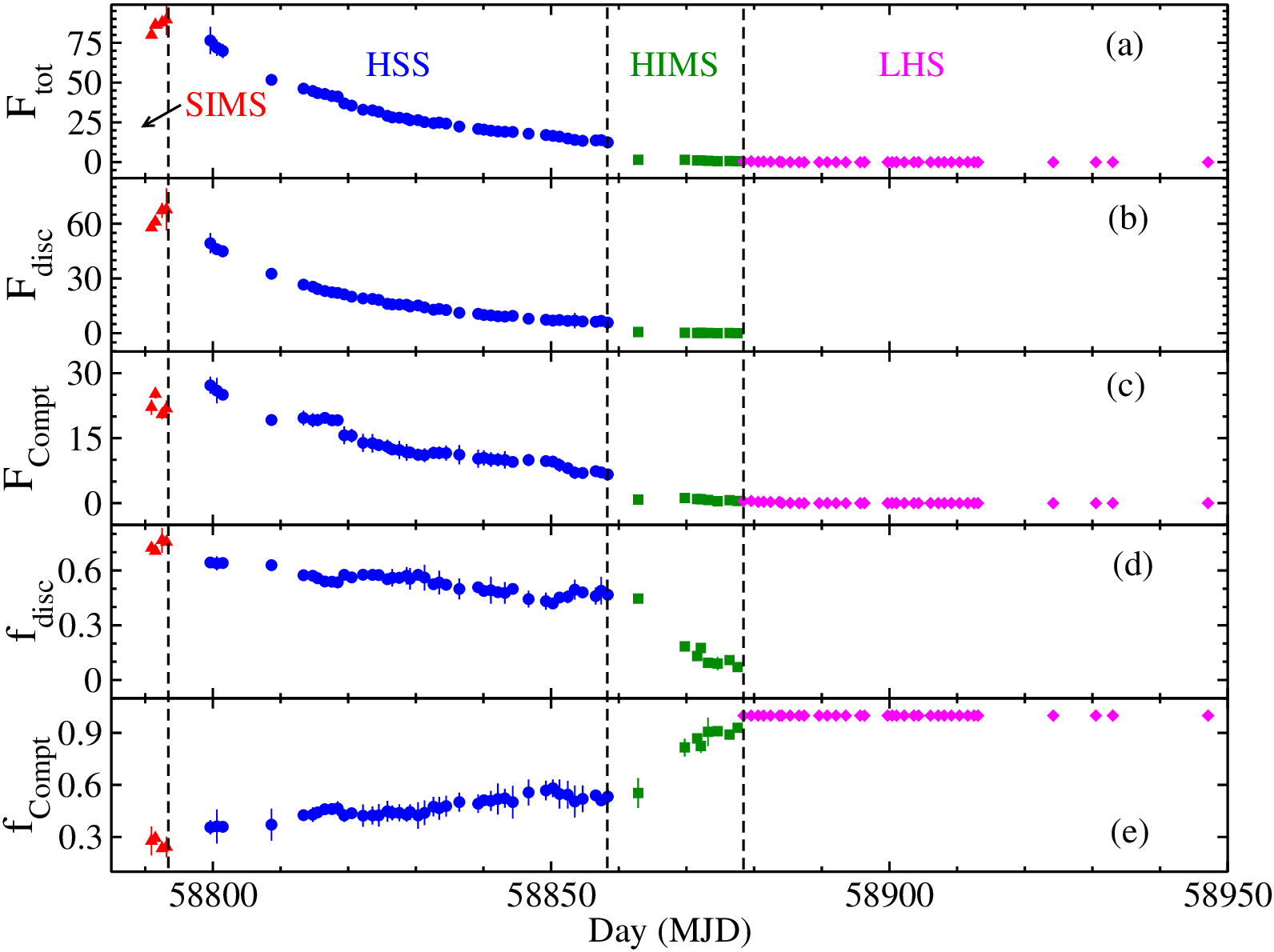}
\caption{The changes in the (a) $0.7-10$~keV unabsorbed flux, (b) $0.7-10$~keV thermal disc flux ($F_{\rm disc}$), (c) $0.7-10$~keV non-thermal Comptonized flux ($F_{\rm Compt}$), (d) thermal fraction ($f_{\rm disc}$),  and (e) non-thermal Comptonized fraction ($f_{\rm Compt}$) are shown. The fluxes are in the unit of 10$^{-10}$ erg cm$^{-2}$ s$^{-1}$. Thermal and non-thermal fractions are defined as $f_{\rm disc} = F_{\rm disc}/F_{\rm tot}$, and $f_{\rm Compt}=F_{\rm Compt}/F_{\rm tot}$, respectively. The colours and symbol correspond to the different spectral states of the source during the outburst as described in Fig.~\ref{fig:lc}. The vertical dashed lines separate different spectral states.}
\label{fig:spec-flux}
\end{figure*}

\begin{figure}
\includegraphics[width=8.5cm]{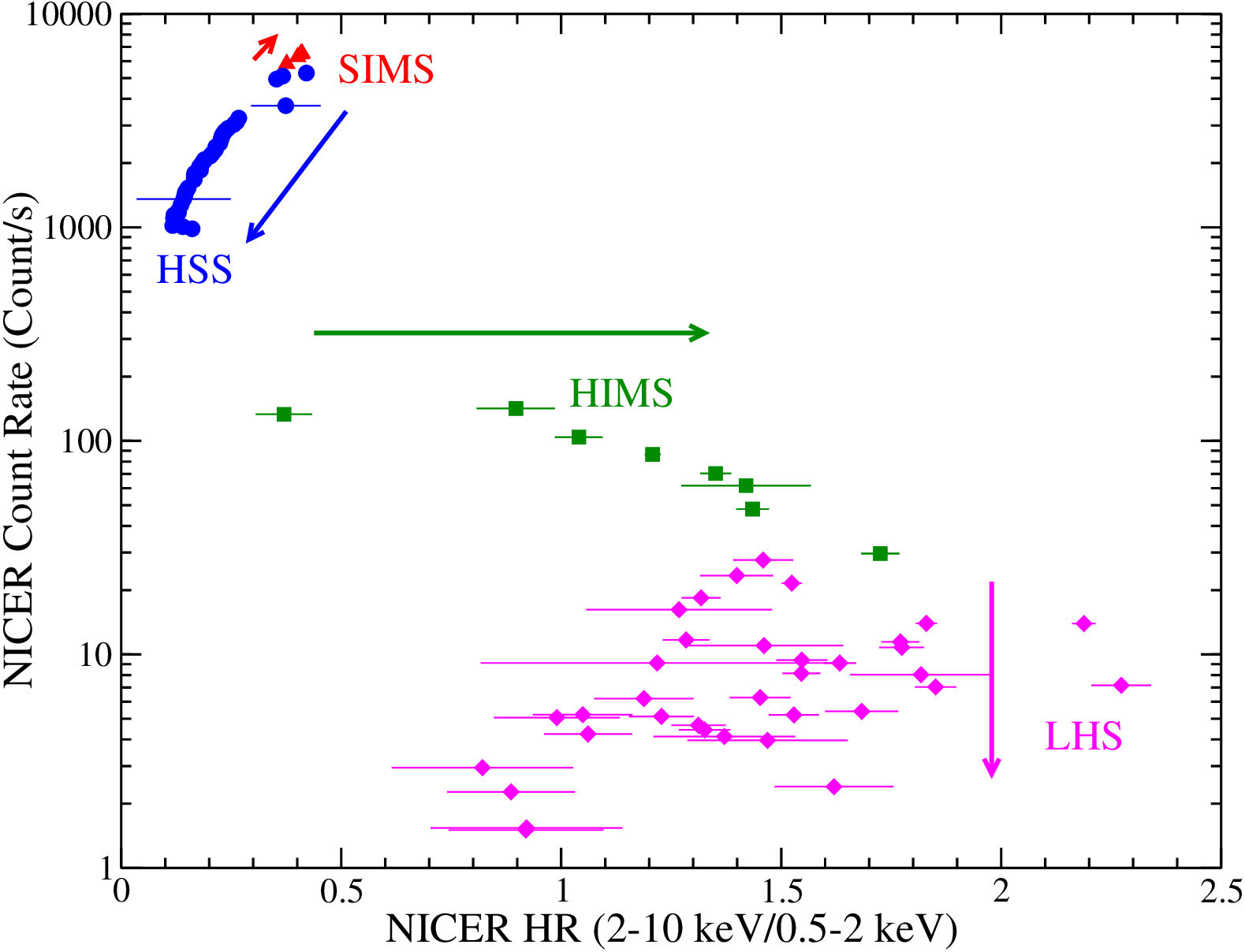}
\caption{The hardness ratio -- intensity diagram (HID) obtained from the $0.5-10$~keV {\it NICER} observations of MAXI~J0637--430 is shown. The colours and symbol correspond to the different spectral states of the source during the outburst as described in Fig.~\ref{fig:lc}.}
\label{fig:hid}
\end{figure}

\begin{figure}
\includegraphics[width=8.5cm]{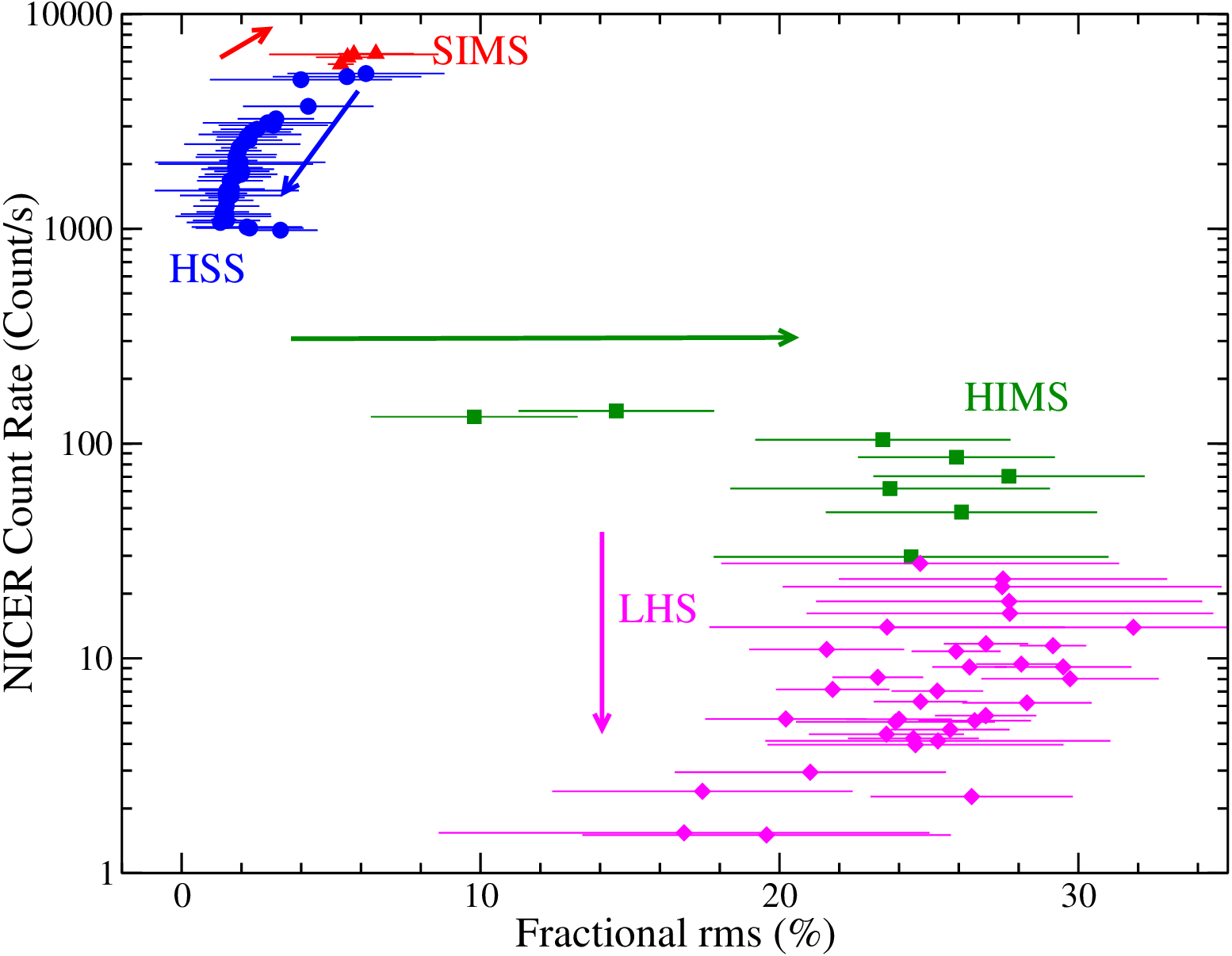}
\caption{The rms--intensity diagram (RID) where the $0.5-10$~keV count rate is plotted as a function of the $0.1-50$~Hz fractional rms amplitude. The colours and symbol correspond to the different spectral states of the source during the outburst as described in Fig.~\ref{fig:lc}.}
\label{fig:rid}
\end{figure}

\subsection{Spectral Properties}
\label{sec:spectral}

We carried out spectral studies of the BHC MAXI~J0637--430 during its 2019-2020 X-ray outburst using the data from {\it NICER} and {\it Swift} observations. A total of 88 {\it NICER} and 85 {\it Swift}/XRT spectra were analyzed in our work. The spectral fitting was carried out in 0.7--10 keV energy band using HEASARC's spectral analysis software package {\tt XSPEC} v12.10\footnote{\url{https://heasarc.gsfc.nasa.gov/docs/xanadu/xspec/}}\citep{Arnaud1996}. The source and the background spectra were extracted by following mission-specific standard procedures, as described in Section~2. Appropriate effective area files and response matrices were used in the spectral fitting. We attempted to fit each spectrum with an absorbed powerlaw model. While fitting, the presence of positive residuals in soft X-ray ranges, specifically in the rising and initial part of the decaying phase of the outburst, allowed us to add a multi-colour blackbody component \citep{Mitsuda1984,Makishima1986} with the absorbed powerlaw model. The model read in {\tt XSPEC} as {\tt tbabs(diskbb+powerlaw)}. In our fitting, the {\tt tbabs} component with {\tt wilm} abundances \citep{Wilms2000} and cross-section of \citet{Verner1996} were used for examining the effect of interstellar absorption. The value of the hydrogen column density ($N_{\rm H}$) was found to vary in the range of $(0.5-1.8) \times 10^{21}$ cm$^{-2}$. As the hydrogen column density in the direction of the source is likely to be constant during the outburst, the observed wide range of variation of $N_{\rm H}$ can be due to the effect of {\tt powerlaw} continuum model spectral fittings. Thus, we replaced the {\tt powerlaw} continuum model with the more robust {\tt nthcomp} model \citep{Zycki1999}. This model read as {\tt tbabs(diskbb+nthcomp)} in {\tt XSPEC}. During fitting, we linked the seed photon temperature ($T_{\rm bb}$) of the {\tt nthcomp} model with the inner disc temperature ($T_{\rm in}$) of the {\tt diskbb} component. We also fixed the Compton corona temperature ($kT_{\rm e}$) at 1000~keV as it is expected to be beyond the {\it NICER} energy range. With this model, we observed that the $N_{\rm H}$ varied in a narrow range of $(1-6) \times 10^{20}$ cm$^{-2}$. Then, we attempted to fit all 88 observations simultaneously with $N_{\rm H}$ linked across the spectra. The simultaneous fitting gave us a good-fit with $\chi^2 = 28805 $ for 25336 dof ($\chi^2_{\rm red}=1.136$, for 88 observations). We obtained $N_{\rm H} = (5.04 \pm 0.52) \times 10^{20}$ cm$^{-2}$ from the simultaneous fitting which is consistent with the Galactic value of $N_{\rm H} = 5.2 \times 10^{21}$ cm$^{-2}$ \citep{HI4PI2006}. Thus, we fixed the hydrogen column density at the Galactic value during our analysis. 

Four representatives \nicer spectra at different epochs of the outburst are shown in the top panel of Fig.~\ref{fig:spec}. The residuals obtained from the spectral fitting with the above model are presented in the bottom panels of Fig.~\ref{fig:spec}. While fitting, we noticed that the thermal disc component was only required in the first 54 \nicer observations. Beyond 29 January 2020 (MJD 58877.58), the multi-colour disc blackbody component was so faint that it was not detected in the spectral fitting. To check whether the {\tt diskbb} component was required in the spectral fitting, we ran {\tt FTOOLS} task {\tt ftest} and found that the {\tt diskbb} component was not required. Nonetheless, the UL on the disc flux in the LHS was estimated to be $\lesssim 10^{-14}$ erg cm$^{-2}$ s$^{-1}$, which is less than 1\% of the Comptonized flux. Therefore, the spectra beyond 29 January 2020 were fitted with an absorbed {\tt nthcomp} model only. Both the models, an absorbed {\tt nthcomp} model with thermal component (for spectra till 29 January 2020) and without thermal component (for spectra beyond 29 January 2020), provided a good fit with the $\chi^2_{red}$ values close to 1. The evolution of various spectral parameters such as (a) the inner disc temperature ($T_{\rm in}$), (b) the {\tt diskbb} normalization ($N_{\rm DBB}$), (c) the photon indices ($\Gamma$), and (d) the {\tt nthcomp} normalization are shown in Fig.~\ref{fig:spec-param}. In the rising phase of the outburst (red triangles), the inner disc temperature ($T_{\rm in}$) was maximum and in the range of $\sim 0.55-0.60$~keV. The inner disc temperature decreased gradually from $\sim 0.6$~keV on 6 November 2019 to $\sim 0.11$~keV on 29 January 2020 (in HSS and HIMS). The photon index ($\Gamma$) in this phase decreased from $5.03$ to $3.23$, though similar variations (in the range of $3.17-4.97$) were seen during the HSS in the declining phase of the outburst (blue circles in Fig.~\ref{fig:spec-param}). During the HIMS and LHS (green squares and magenta diamonds in Fig~\ref{fig:spec-param}, respectively), the spectra became flat with $\Gamma$ in the range of $\sim 1.68-2.62$. The {\tt diskbb} normalization ($N_{\rm DBB}$) varied in the range of $\sim 2700-3500$ in the SIMS and HSS. However, in the HIMS, we observed a higher value of $N_{\rm DBB}$ ($>8000$). 

Interestingly, we did not find any signature of the presence of iron emission line in the spectral residuals. Thus, we attempted to estimate the UL on the equivalent width (EW) of the Fe K$\alpha$ emission line at 6.4 keV. In the beginning, we added a narrow Gaussian line with fixed-line energy (at 6.2, 6.4, 6.5 and 6.7 keV) and line width (0.01, 0.05, 0.1, and 0.2 keV) to calculate the EW. We observed that if an iron line was present, it is most likely at $\sim 6.4$~keV, where the maximum value of EW was obtained for a line width of 0.2 keV. The UL on the EW of the iron emission line was 0.019 keV, 0.067 keV, 0.007 keV and $<10^{-4}$~keV for SIMS, HSS, HIMS and LHS, respectively. While fitting the averaged spectra from the four spectral states, the UL on EW was estimated to be 0.020 keV, 0.091 keV, 0.031 keV and 0.002 keV, respectively.
 
Unabsorbed flux in $0.7-10$~keV range ($F_{\rm tot}$), estimated from the spectral fitting of the {\it NICER} observations, are shown in Fig.~\ref{fig:spec-flux}(a). Thermal ($F_{\rm disc}$) and non-thermal Comptonized flux ($F_{\rm Compt}$) are shown in panels (b) and (c) of Fig.~\ref{fig:spec-flux}, respectively. In Fig.~\ref{fig:spec-flux}(d) \& (e), the thermal fraction ($f_{\rm disc}$), and the non-thermal Comptonized fraction ($f_{\rm Compt}$) are shown. The thermal and non-thermal fractions are defined as the ratio between $F_{\rm disc}$ \& $F_{\rm tot}$, and $F_{\rm Compt}$ \& $F_{\rm tot}$, respectively. In the rising phase of the outburst, the $0.7-10$~keV unabsorbed total flux as well as the thermal multi-colour disk blackbody flux ($F_{\rm disc}$) were increasing. The $0.7-10$~keV unabsorbed source flux became maximum ($F_{\rm tot} = 8.96 \times 10^{-9}$ erg cm$^{-2}$ s$^{-1}$) at the peak of the outburst on 6 November 2019 (MJD 58793.14). In the declining phase, we observed a monotonous decrease in $F_{\rm tot}$ and $F_{\rm disc}$. Though the non-thermal Comptonized flux ($F_{\rm Compt}$) also decreased during the declining phase, the variation was not as smooth as $F_{\rm tot}$ and $F_{\rm Compt}$. In the SIMS and the HSS (red \& blue data points in the figure, respectively), the source flux in $0.7-10$~keV energy band was dominated by the thermal disc flux. Particularly in the rising phase, the thermal fraction was very high at $f_{\rm disc} > 0.7$ and decreased slowly to $f_{\rm disc} \geq 0.6$ as the outburst progressed (HSS). In the HIMS, the Comptonized flux started to dominate with $f_{\rm Compt} \geq 0.5$. In the LHS, however, only non-thermal Comptonized flux ($f_{\rm Compt}$) was observed due to non-detection of thermal component. The best-fit parameters obtained from our spectral analysis are given in Table~\ref{tab:analysis}.

We attempted to fit the spectra by replacing {\tt diskbb} model component with {\tt diskpn} model \citep{Gierlinski1999}. The {\tt diskpn} component is an extension of {\tt diskbb} model by incorporating corrections for the temperature distribution near the black hole by considering the torque-free inner-boundary condition. Results obtained by fitting data with absorbed {\tt diskpn+nthcomp} model are found to be comparable to that obtained for absorbed {\tt diskb+nthcomp} model. The variation of the maximum disc temperature ($T_{\rm max}$) obtained from the {\tt diskpn} model are same as the inner disc temperature ($T_{\rm in}$), obtained from the {\tt diskbb} model, within uncertainties.

We also fitted the $0.5-10$~keV {\it Swift}/XRT spectra from all 85 epochs of observation between 7 November 2019 and 30 March 2020 using the above model. The results obtained from the fitting are broadly similar to that obtained from the {\it NICER} observations, though the errors estimated from the fitting of {\it Swift}/XRT data were large. Therefore, we did not describe here in detail.

\subsection{Evolution of Spectro-Temporal Properties}
\label{sec:spec-temp}

As described in the previous section, we classified the entire outburst into four different spectral states such as SIMS (rising phase - red triangles), HSS (declining phase - blue circles), HIMS (declining phase - green squares), and LHS (declining phase - magenta diamonds). The evolution of the source among these spectral states can also be recognized while studying the correlation between the observed spectral and timing properties. For this purpose, we generated the hardness-intensity diagram (HID) and rms-intensity diagram (RID). In Fig.~\ref{fig:hid} and Fig.~\ref{fig:rid}, the HR (ratio between the count rates in $2-10$~keV and $0.5-2$~keV ranges) and the $0.01-50$~Hz fractional rms amplitude are plotted as a function of the source count rate in $0.5-10$~keV energy range, respectively. 

From Fig.~\ref{fig:hid} \& Fig.~\ref{fig:rid}, we observed that the HR and rms were positively correlated with the source count rate during the SIMS in the rising phase (red triangles). In this phase of the outburst, the track of the source in the HID and RID was from left to right, as marked with red arrows in both the figures. In the HSS, the source intensity decreased gradually along with the decrease in HR and fractional rms amplitude values. The evolution track of the source in the HID and RID were in the opposite direction to that during the rising phase (SIMS) of the outburst (blue arrows in Fig.~\ref{fig:hid} \& Fig.~\ref{fig:rid}). This trend was clearly observed until 10 January 2020. Beyond 10 January 2020, the evolution track of the source was changed from left to right in the HID and RID (see Fig.~\ref{fig:hid} \& Fig.~\ref{fig:rid}). The HR and fractional rms amplitude started increasing rapidly with the decrease in source count rate as the source were in the HIMS (see Fig.~\ref{fig:hid} \& Fig.~\ref{fig:rid}). After 29 January 2020, the HR and fractional rms amplitude varied randomly as the count rate decreased further (see Fig.~\ref{fig:hid} \& Fig.~\ref{fig:rid}) when the source was in the LHS.

\begin{figure}
\includegraphics[width=8.5cm]{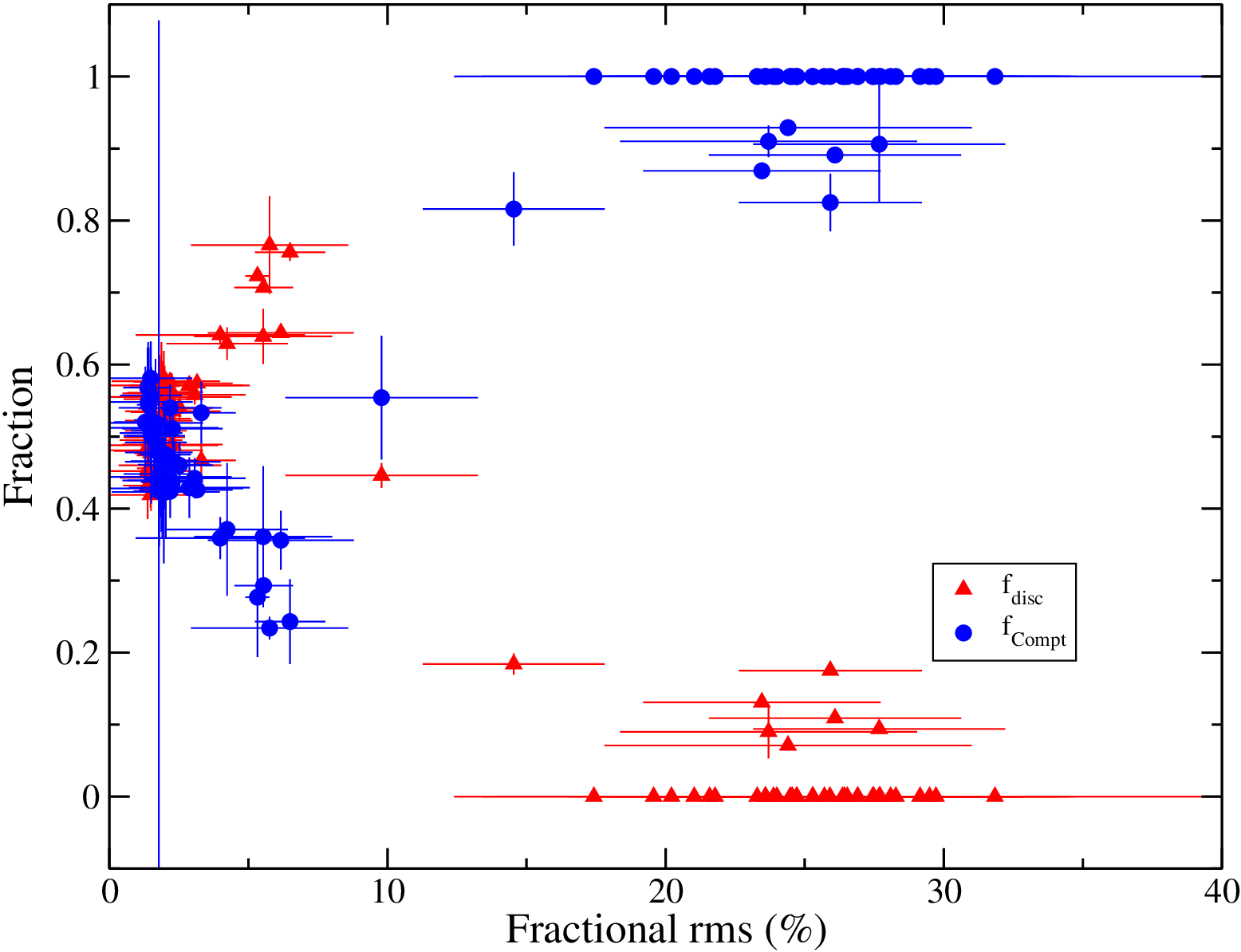}
\caption{The thermal (disc) and non-thermal (Comptonized) emission fractions are plotted as a function of fractional rms amplitude ($r$). Red triangles and blue circles represent the thermal ($f_{\rm disc}$) and Comptonized flux fraction ($f_{\rm Compt}$), respectively.}
\label{fig:rms}
\end{figure}

\section{Discussion}
\label{sec:discussion}

We present a detailed analysis of the spectral and timing properties of the 2019-2020 outburst of the recently discovered BHC MAXI~J0637--430. We used a total of 88 epochs of \nicer observations in $0.7-10$~keV energy band and 85 observations with the {\it Swift}/XRT in the energy band of $0.5-10$~keV. The spectra were fitted with a thermal blackbody and non-thermal Comptonized model. Among the total of 88 \nicer observations, the thermal disc component was required in the first 54 observations. We found that the source evolved through SIMS $\rightarrow$ HSS $\rightarrow$ HIMS $\rightarrow$ LHS during the entire outburst.

\subsection{Nature of the X-ray source}
\label{sec:nature}

The nature of the compact object in MAXI~J0637--430, though believed as a black hole, has not been confirmed yet. Based on the results obtained from a {\it NuSTAR} observation in $3-79$ keV range on 5 November 2019, \cite{Tomsick19} suggested that the system is a BHXRB. As the mass of the compact object has not yet been estimated through the dynamical process and lack of clear signature of neutron star (NS) system such as pulsations and/or thermonuclear bursts, the nature of the compact object is unclear. Thus, we attempted to infer the nature of the compact object from the timing and spectral properties. We searched for the presence of high-frequency kHz QPOs, seen in several NSs \citep{disalvo2001,vanDoesburgh2017,Bult2018,Mendez2021}, in the PDS up to 1250 Hz. However, we did not find any signs of such features in the PDS. Absence of kHz QPOs, however, did not discard the compact object as an NS. The PDS showed that the power decreased rapidly in frequency $>10$~Hz, which is consistent with the BHXRB \citep{Sunyave2000}. Generally, the power diminishes in the frequency range of $\sim 10-50$~Hz for BHXRB, while the NSXRBs show variability up to as high as $\sim100-200$~Hz. The evolution of HID and RID of MAXI~J0637--430 is similar to that of other BHXRB \citep{Homan2005,Belloni2005,RM06,Alabarta2020,Zhang2020}. However, several NS binary systems, especially at low accretion rate, also follow a similar evolution track \citep{Kording2008,Munoz-Darias2011,Munoz-Darias2014}. The state transition track in low accreting NS is diagonal, while it is horizontal in the case of BH \citep{Munoz-Darias2014}. Though {\em RXTE} and \nicer operate in different energy ranges with certain overlap, for a given source, the HID and RID from both sets of instruments are similar \citep{Alabarta2020}. While comparing the evolution tracks of HID and RID of BHCs GX~339--4, MAXI~J1727--203, MAXI~J1348--630 \citep{Belloni2005,Munoz-Darias2014,Alabarta2020,Zhang2020} and MAXI~J0637--430 (present work, see Fig.~\ref{fig:hid} and Fig.~\ref{fig:rid}), it can be seen that the state transition track followed a horizontal path in case of all the sources. Similar evolution track of HID and RID as other BHCs and non-detection of any characteristics of neutrons stars such as pulsations, thermonuclear bursts and kHz QPOs in the PDS suggest that the X-ray source in the MAXI~J0637--430 binary system is likely to be a BH.

The evolution of the spectral properties of MAXI~J0637--430 is consistent with that of BH LMXB. Usually, the BH spectra contain an ultra-soft multi-colour blackbody component ($T \lesssim 1$~keV) and a power-law tail. However, the NS spectra contain a soft component that originates from the surface of the NS and a multi-colour disc blackbody component from the accretion disc along with Comptonized power-law tail \citep{Fu1990}. In this work, the $0.7-10$~keV {\it NICER} spectra were fitted with a ultra-soft thermal component ($T_{\rm in} \lesssim 0.6$~keV) and a power-law tail (see, Section~\ref{sec:spectral}), considering the compact object as a BH. In order to check the NS nature of the X-ray source in the binary, we carried out the spectral fitting of the data taken during the bright phase of the outburst with a more general NS model consisting of emission from the surface of the NS ({\tt bbodyrad}), accretion disc ({\tt diskbb}) along with the power-law continuum model ({\tt nthcomp}). Parameters corresponding to temperatures of the NS surface and accretion disc, obtained from fitting the data with this NS specific model, are $kT_{\rm bbodyrad} \sim 0.6\pm0.2 -0.1\pm0.1$~keV and $T_{\rm in} \sim 0.6\pm0.01 -0.1\pm0.005$~keV, respectively. The normalization of the {\tt bbodyrad} component was very low ($\sim 2.5\pm2$) compared to that of {\tt diskbb} component ($> 2500$). Although the fitting was acceptable with a reduced $\chi^2$ $\sim$ 1, it was clear that the thermal component corresponding to the emission from the NS surface was insignificant and not required. We also used {\tt F-test} statistic to check if the second thermal component was required in the spectral fitting. The {\tt F-test} rejected the probability of two thermal components in the spectra. Observed timing and spectral characteristics of MAXI~J0637--430; therefore, suggest that the compact object is a black hole.

\begin{figure}
\includegraphics[width=8.5cm]{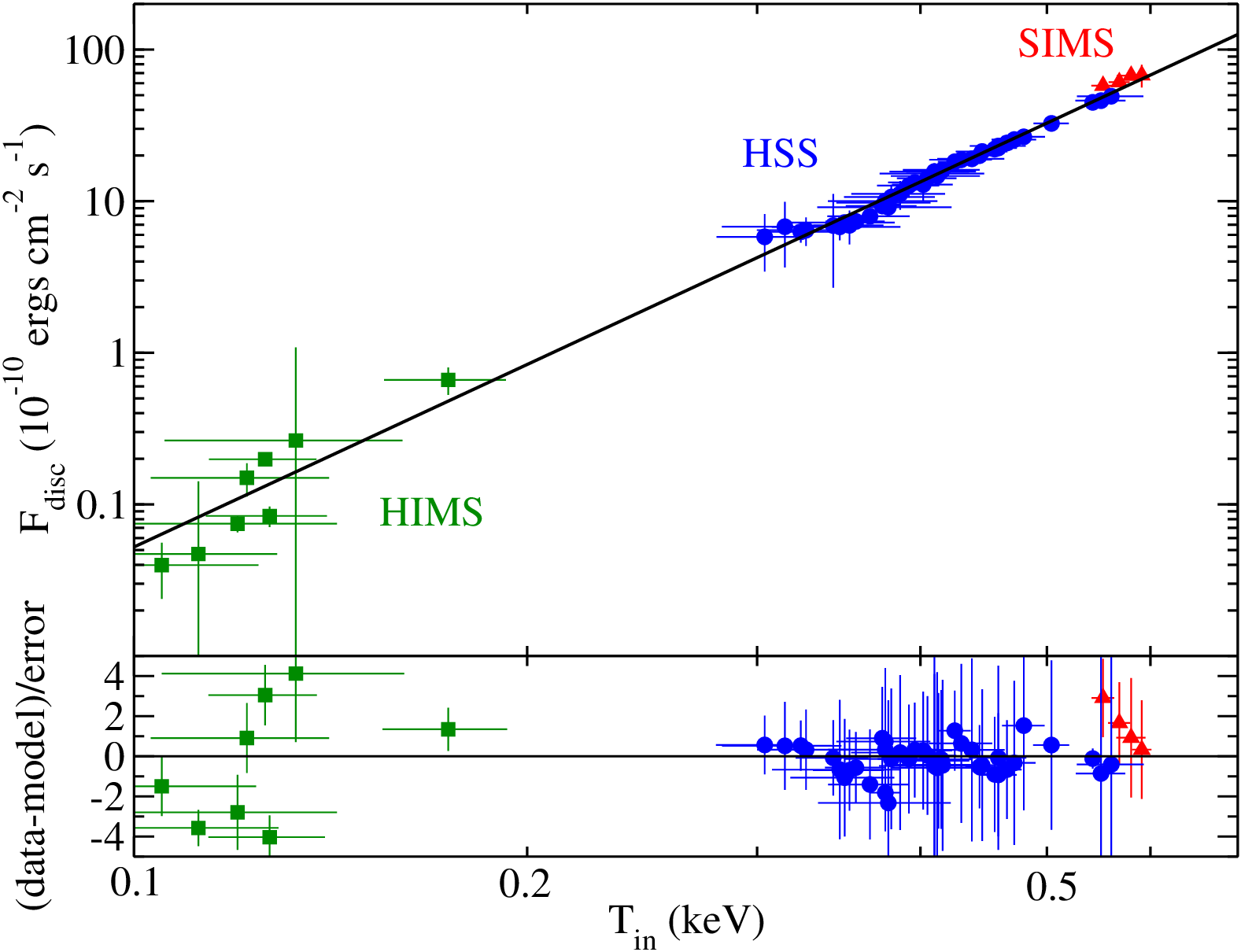}
\caption{Disk blackbody flux ($F_{\rm disc}$) is plotted as a function of inner disc temperature ($T_{\rm in}$) in the upper panel. Solid black line represent $F_{\rm disc} \sim T_{\rm in}^{\rm 4}$ line. In the lower panel, residual is plotted. The colours and symbol correspond to the different spectral states of the source during the outburst as described in Fig.~\ref{fig:lc}.}
\label{fig:f-t4}
\end{figure}

\subsection{Evolution of the Outburst}
\label{sec:evo}

An outburst is thought to be triggered by the sudden enhancement of the viscosity at the outer edge of the disc \citep{Ebisawa1996}. In the initial phase of the outburst, LHS is observed with a disc truncated at a large distance. In this state, the Comptonized emission dominates the spectrum with $F_{\rm Compt}>80 $\% \citep{RM06}. As the outburst progresses, the disc emission starts to dominate, and the source moves through the HIMS, SIMS and HSS. The disc is observed to move closer to the BH in the rising phase of the outburst. The source entered the declining phase when the viscosity is turned off, and the matter supply is cut-off. As the disc is already formed, it is not easy to accrete the disc matter in the absence of viscosity \citep{Roy2017}. The disc gets accreted slowly to the BH; hence we see relatively long HSS and SIMS in the declining phase. Once the disc matter is accreted on to the BH, the source quickly evolves through the HIMS and LHS, as the supply of the matter has already been cut-off. This is why very short HIMS and LHS are observed in the declining phase of an X-ray outburst.

In the 2019-2020 outburst of MAXI~J0637--430, the LHS and the HIMS were not observed. It is possible that the source moved quickly through the harder states (LHS and HIMS). However, this transition could not be investigated due to the lack of X-ray observations during this phase. The source was caught in SIMS with \nicer only $\sim 4$ days prior to the peak of the outburst. After reaching the peak, the source entered the declining phase during which the X-ray intensity slowly decreased over the next $\sim 65$ days, which implies that nearly two months duration was required to drain the disc. As the matter supply was already cut-off and very little matter remained, the X-ray intensity suddenly dropped by $\sim 10$ times once the disc matter was drained. Then, the source quickly evolved through the HIMS and entered to a very low luminosity LHS.

Evolving type-C QPOs are known to be detected in the HIMS and LHS, whereas the sporadic type-A or type-B QPOs may be observed in the SIMS. However, no QPOs are observed in the HSS \citep{Belloni2005,RM06}. The oscillation of the Compton cloud is believed to be responsible for the QPOs \citep{MSC1996,TLM1998,CM2000,Cabanac2010}. It is well established that the Comptonized photons are responsible for the variability observed in the PDS and light curves \citep{vanderklis2004}. In general, strong variability is observed in the LHS with fractional rms amplitude of $\sim 20-40 \%$. In the HIMS and SIMS, fractional rms amplitudes of $10-20\%$ and $\sim 5-10$ \% are seen, respectively. However, in the HSS, weak variability with fractional rms amplitude of $<5\%$ are observed \citep{vanderklis1994,Mendez1997,Belloni2005}. In the current outburst of MAXI~J0637--430, weak variability is observed in the SIMS and HSS with fractional rms $r \leq 5 \%$ (red and blue colour in Fig.~\ref{fig:lc}). Strong variability are observed in the LHS and HIMS with $r > 20 \%$ and $\sim 10-20 \%$ (magenta and green colors in Fig.~\ref{fig:lc}), respectively, as the Comptonized hard photons dominate these states. In Fig~\ref{fig:rms}, we plot fractional rms amplitude as a function of the thermal and non-thermal fraction. From this figure, it is clear that strong variability is observed when the non-thermal fraction is higher, i.e., Comptonized photons dominate the total flux.

We tried to find the dependency of disc flux ($F_{\rm disc}$) with the inner disc temperature ($T_{\rm in}$). We fitted the distribution with a relation $F_{\rm disc} \sim T_{\rm in}^{\rm b}$. The best-fit exponent is derived to be $b=4.24\pm 0.14$. This is higher than the theoretical value of $b=4$. In the upper panel of Fig.~\ref{fig:f-t4}, we show thermal disc flux ($F_{\rm disc}$) as a function of inner disc temperature ($T_{\rm in}$). The disc flux was fitted with $F_{\rm disc}\sim T_{\rm in}^{\rm 4}$ relation, and the residual is plotted in the lower panel of  Fig.~\ref{fig:f-t4}. The red triangles, blue circles and green diamonds represent data from the SIMS, HSS and HIMS, respectively. The solid black line represents the theoretical relation, $F_{\rm disc} \sim T_{\rm in}^{\rm 4}$. From Fig.~\ref{fig:f-t4}, it can be seen that the observed disc flux was in agreement with the theoretical value in the HSS and SIMS. However, there was a deviation from the theoretical value in the HIMS when the disc flux was low. The $F_{\rm disc} \sim T_{\rm in}^{\rm 4}$ relation is predicted for a non-rotating black hole, assuming the inner edge and spectral hardening factor to remain constant \citep{Gierlinski2004}. Therefore, the deviation in the HIMS could be due to the moving inner edge or change in the hardening factor or both \citep{Dunn2011}. Similar behaviour is also observed for other black hole X-ray binaries \citep{Gierlinski2004,Dunn2010,Dunn2011}.

In our analysis, we did not detect any signature of the presence of the Fe K$\alpha$ emission line, which is very unusual. Thus, we calculated the UL on the EW of the Fe K$\alpha$ line in the \nicer spectra. The UL was estimated to be 0.019 keV, 0.067 keV, 0.007 keV and $<10^{-4}$~keV in the SIMS, HSS, HIMS and LHS, respectively. The averaged spectra from each of the four spectral states gives UL on the EW as 0.020 keV, 0.091 keV, 0.011 keV and 0.002 keV, respectively. From this, it is observed that a weak Fe line might be present in the spectra of HSS. Non-detection of Fe emission line could also be due to the short exposure time of \nicer observations. In the case of 2018 outburst of MAXI~J1813-095, the Fe K$\alpha$ line was not detected with the {\it NICER} observation with short exposure time ($\sim 1-2$ ks), but \nustar observation with a long exposure ($\sim 20$ ks) detected a broad iron line \citep{AJ2021}. Indeed in the current outburst, \cite{Tomsick19} observed an Fe K$\alpha$ line with the \nustar observation, which has long exposure. We also did not observe reflection hump in the spectra as it is expected in $\sim 15-40$~keV range \citep{Fabian1989,Matt1991}, which is outside of \nicer energy range.

\begin{figure}
\includegraphics[width=8.5cm]{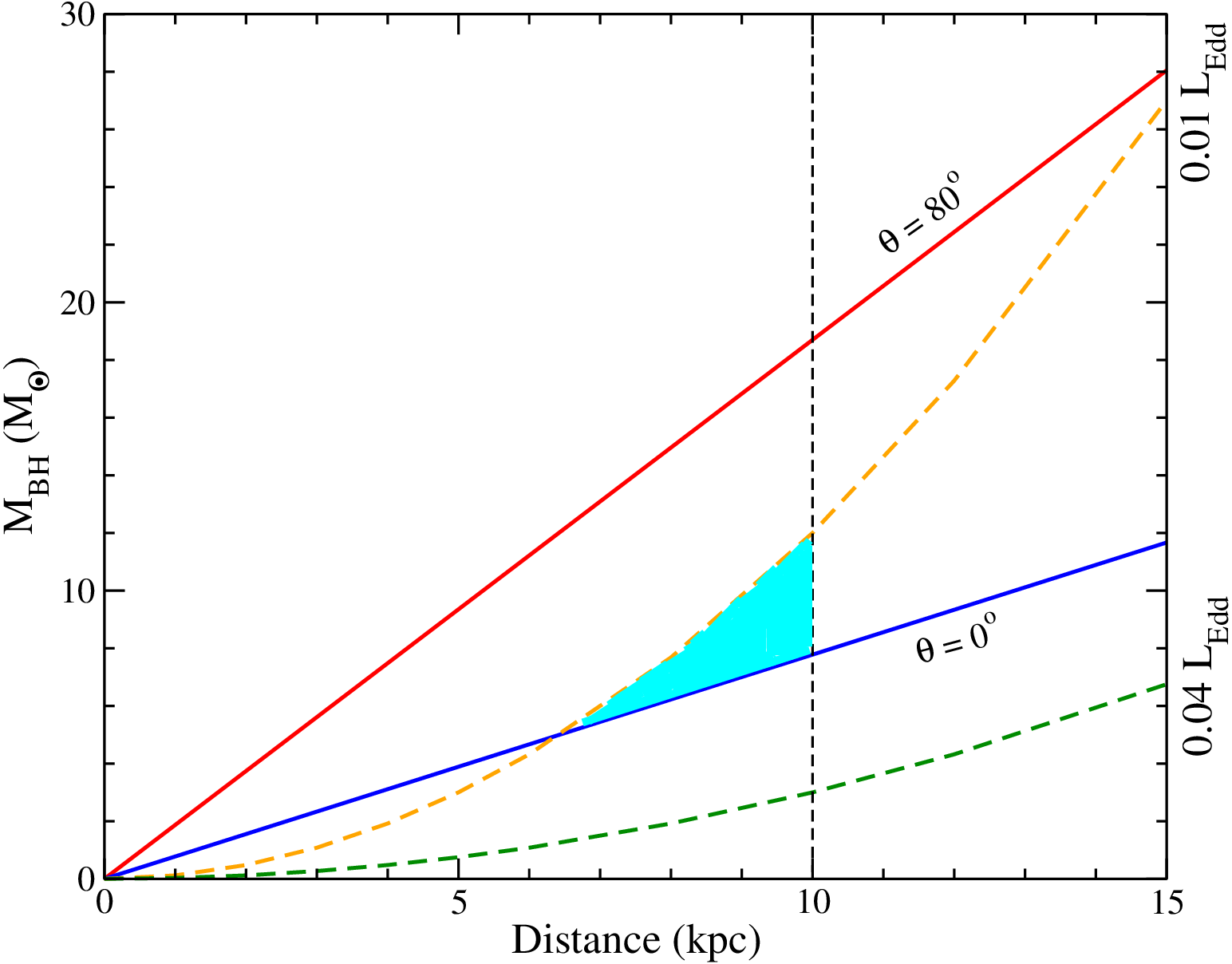}
\caption{Mass-distance relation for MAXI~J0637--430 is plotted assuming a non-rotating black hole. Red and blue solid lines correspond to the inclination angle $\theta=80$\textdegree and $\theta=0$\textdegree, respectively. The orange dashed and green dashed curves show the mass-distance relation for $L_{\rm tr}/L_{\rm Edd} = 0.01$ and $L_{\rm tr}/L_{\rm Edd} = 0.04$, respectively. The shaded region is the common parameter space of $M_{\rm BH}-d$, obtained from Equation~\ref{eqn:2} and Equation~\ref{eqn:3}. The vertical dashed black line represent $d=10$~kpc.}
\label{fig:mass}
\end{figure}

\subsection{Estimation of mass and distance of MAXI~J0637--430}
\label{sec:mass-dist}

In the softer spectral states (SIMS and HSS), the accretion flow is expected to be truncated at very close to the BH. During the SIMS and HSS of the outburst, the normalization ($N_{\rm DBB}$) of the {\tt diskbb} model did not vary much. It varied in the range of $\sim 2700-3500$ indicating that the inner edge ($R_{\rm in}$) of the disc was remarkably stable in the SIMS and HSS \citep{Ebisawa1993}. We tried to constrain the mass and distance of the BH by equating $R_{\rm in}$ with the innermost stable circular orbit (ISCO). The normalization of the {\tt Diskbb} model is given by, $N_{\rm DBB} = (r_{\rm in}/D_{\rm 10})^2 \cos \theta$, where $r_{\rm in}$ is apparent disc radius in km, $D_{\rm 10}$ is the source distance in the unit of 10 kpc, and $\theta$ is the inclination angle. The inner edge of the disc ($R_{\rm in}$) is given by, $R_{\rm in}=\xi \kappa^2 r_{\rm in}$, where $\xi = 0.41$ is correction factor \citep{Kubota1998}, $\kappa = 1.7-2.0$ is the spectral hardening factor \citep{Shimura-Takahara1995}. Using, $\kappa = 1.8$, we get,

\begin{equation}
\label{eqn:1}
R_{\rm in} = (73 \pm 4) D_{10} (\cos \theta)^{-1/2}. 
\end{equation}

In our calculation, we used the mean value of $N_{\rm DBB}=3045 \pm 175$ from corresponding observed values during the SIMS and HSS. For a Schwarzschild black hole, $R_{\rm ISCO} = 6 r_g$, where $r_g$ is gravitational radius. From these equations and assuming a non-rotating BH, we obtained the mass of the BH as, 

\begin{equation}
\label{eqn:2}
M_{\rm BH} = (8\pm1)D_{\rm 10}(\cos \theta)^{-1/2} M_{\sun}.
\end{equation}

We plotted the mass-distance relation for MAXI~J0637--430 in Fig.~\ref{fig:mass} assuming a non-rotating black hole, using Equation~\ref{eqn:2}. Red and blue solid lines correspond to the inclination angle $\theta=80$\textdegree and $\theta=0$\textdegree, respectively. From this, the $M_{\rm BH}$ should lie in the parameter space between these two line. However, $M_{\rm BH}$ is not well-constrained and depends on the source distance. Assuming the source distance, $d<10$ kpc, the value of the mass of the BH would be $M_{\rm BH}<18$ $M_{\sun}$. The above calculation may not hold for a rotating BH, since, the {\tt diskbb} model assumes a non-rotating BH.

To constrain the mass and distance of the BH better, we used spectral state transition luminosity relation \citep{Maccarone2003,Tetarenko2016}. Empirically, soft-to-hard state transition luminosity ($L_{\rm tr}$) is observed as,

\begin{equation}
\label{eqn:3}
L_{\rm tr} \sim 0.01-0.04~L_{\rm Edd}.
\end{equation}

Recently, \cite{AJ2020b} estimated the distance of MAXI~J1348--630 using this relation. We calculated the state-transition luminosity ($L_{\rm tr}$) from the unabsorbed flux obtained on 10 January 2020. In Fig.~\ref{fig:mass}, the orange dashed and green dashed curves show the mass-distance relation for $L_{\rm tr}/L_{\rm Edd} = 0.01$ and $L_{\rm tr}/L_{\rm Edd} = 0.04$, respectively. The shaded region in Fig~\ref{fig:mass} shows the common parameter space of $M_{\rm BH}-d$ obtained from Equation~\ref{eqn:2} and Equation~\ref{eqn:3}. From this, we infer that the mass of the BH is in the range of $5-12$ $M_{\sun}$, for the source distance of $d<10$~kpc. This mass-distance relation also infers that the source is located at a distance of $d>6.5$~kpc (lower end of the shaded region in Fig.~\ref{fig:mass}). However, our estimation of the mass, distance and inclination angle is based on the assumption of a non-rotating black hole. A more detailed optical study is required to estimate the distance and/or inclination angle of the source.

\section{Summary}
\label{sec:summary}
We studied the 2019-2020 outburst of MAXI~J0637--430 with \nicer and {\it Swift} observations. Following are the key findings from this study.

\begin{enumerate}

\item From the detailed timing and spectral studies, we conclude that the compact object in MAXI~J0637--430 is likely to be a black hole.  

\item We classified the outburst in four spectral states: SIMS, HSS, HIMS and LHS. The classification was done based on the evolution of the soft X-ray flux, hard X-ray flux, HR, HID and RID.

\item No QPOs were observed during the outburst.
   
\item Weak variability (fractional variability amplitude, $r<5 \%$) are observed in the SIMS and HSS. In the HIMS and LHS, however, strong variability were observed with $r >20 \%$. Comptonized photons are responsible for the observed variability in the PDS and light curves.

\item The $0.7-10$~keV {\it NICER} spectra were fitted with a composite model consisting of disk blackbody and nthcomp, along with the interstellar absorption. Spectra obtained from 54 observations from the beginning when the source was in SIMS, HSS and HIMS, fitted well with this model. However, the thermal disc component was undetectable in the spectral fitting in the LHS.

\item Thermal disc flux and inner disc temperature followed the theoretical relation of $F_{\rm disc} \sim T_{\rm in}^4$ in the HSS and SIMS. A deviation of the relation is observed in the HIMS. Moving inner edge or change of spectral hardening factor could be responsible for that.

\item Assuming a non-rotating BH, we estimated the mass of the BH as $5-12$ $M_{\sun}$ for a distance of $d<10$~kpc.

\item No signature of Fe emission line was seen. This could be due to the short exposure time of \nicer observations.
    
\end{enumerate}

\section*{Acknowledgements}
We acknowledge the reviewer for his/her helpful comments/suggestions which improved the paper. We sincerely thank Jon M. Miller and Zaven Arzoumanian for their constructive suggestions on the paper. This research has made use of data and/or software provided by the High Energy Astrophysics Science Archive Research Center (HEASARC), which is a service of the Astrophysics Science Division at NASA/GSFC and the High Energy Astrophysics Division of the Smithsonian Astrophysical Observatory. This work was made use of XRT data supplied by the UK Swift Science Data Centre at the University of Leicester, UK. Work at Physical Research Laboratory, Ahmedabad, is funded by the Department of Space, Government of India.

\section*{DATA AVAILABILITY}
We used archival data of {\it Swift}, and {\it NICER} observatories for this work.

\bibliographystyle{mnras}
\bibliography{reference}



\appendix

\begin{table*}
\caption{Best-fit parameters obtained from spectral fitting of data obtained from the {\it NICER} observations of MAXI~J0637--430 during its 2019-2020 outburst}
\label{tab:analysis}
\begin{tabular}{lccccccccccccc}
\hline
Obs ID & UT Date & Exp & Date & Count Rate & DBB Flux$^a$ & PL Flux$^a$ & $T_{\rm in}$ & $N_{\rm DBB}$ & $\Gamma$ & $\chi^2$/dof \\
 &(yyyy-mm-dd) &(s) &(MJD) & (Count s$^{-1}$) &  &  &  (keV) & & &  \\
\hline
2200950101 &2019-11-03 & 1080 &  58790.92 &  5843 &$ 57.82\pm{0.71}$&$ 22.13\pm{1.75}$&$ 0.55\pm{0.02}$&$ 2969\pm {88  }$&$ 5.03\pm{0.07}$&  540/ 519 \\ 
2200950102 &2019-11-04 & 6698 &  58791.50 &  6288 &$ 61.06\pm{1.15}$&$ 25.20\pm{1.02}$&$ 0.57\pm{0.01}$&$ 3048\pm {115 }$&$ 4.67\pm{0.05}$&  718/ 696 \\
2200950103 &2019-11-05 & 571  &  58792.50 &  6486 &$ 67.29\pm{1.15}$&$ 20.50\pm{1.16}$&$ 0.58\pm{0.01}$&$ 2973\pm {102 }$&$ 4.71\pm{0.09}$&  512/ 497 \\
2200950104 &2019-11-06 & 334  &  58793.14 &  6533 &$ 67.77\pm{2.25}$&$ 21.83\pm{1.83}$&$ 0.60\pm{0.01}$&$ 3032\pm {93  }$&$ 3.53\pm{0.04}$&  528/ 464 \\
2200950107 &2019-11-12 & 591  &  58799.63 &  5280 &$ 49.27\pm{3.53}$&$ 27.19\pm{1.96}$&$ 0.56\pm{0.03}$&$ 3082\pm {106 }$&$ 3.23\pm{0.08}$&  509/ 538 \\
2200950108 &2019-11-13 & 1821 &  58800.60 &  5108 &$ 46.02\pm{2.26}$&$ 25.96\pm{2.94}$&$ 0.55\pm{0.02}$&$ 2991\pm {134 }$&$ 3.86\pm{0.06}$&  415/ 409 \\
2200950109 &2019-11-14 & 3760 &  58801.47 &  4946 &$ 44.86\pm{3.15}$&$ 25.03\pm{1.07}$&$ 0.54\pm{0.04}$&$ 2801\pm {152 }$&$ 4.50\pm{0.07}$&  521/ 558 \\
2200950111 &2019-11-21 & 18   &  58808.66 &  3715 &$ 32.59\pm{2.18}$&$ 19.18\pm{0.92}$&$ 0.50\pm{0.01}$&$ 2954\pm {244 }$&$ 3.17\pm{0.04}$&  589/ 568 \\
2200950112 &2019-11-26 & 6736 &  58813.39 &  3252 &$ 26.58\pm{0.80}$&$ 19.65\pm{1.71}$&$ 0.48\pm{0.02}$&$ 3160\pm {122 }$&$ 4.06\pm{0.04}$&  613/ 622 \\
2200950113 &2019-11-27 & 1213 &  58814.77 &  3113 &$ 25.51\pm{1.52}$&$ 19.16\pm{1.58}$&$ 0.47\pm{0.02}$&$ 3321\pm {123 }$&$ 3.84\pm{0.06}$&  494/ 471 \\
2200950114 &2019-11-28 & 1936 &  58815.48 &  3034 &$ 24.22\pm{0.74}$&$ 19.17\pm{1.36}$&$ 0.47\pm{0.01}$&$ 3044\pm {423 }$&$ 4.26\pm{0.07}$&  483/ 496 \\
2200950115 &2019-11-29 & 1962 &  58816.55 &  2906 &$ 23.13\pm{1.67}$&$ 19.67\pm{1.30}$&$ 0.46\pm{0.02}$&$ 2774\pm {259 }$&$ 4.85\pm{0.07}$&  519/ 481 \\
2200950116 &2019-11-30 & 2274 &  58817.61 &  2823 &$ 22.44\pm{0.88}$&$ 19.13\pm{1.23}$&$ 0.46\pm{0.01}$&$ 2917\pm {231 }$&$ 4.79\pm{0.06}$&  515/ 478 \\
2200950117 &2019-12-01 & 1778 &  58818.45 &  2748 &$ 22.08\pm{0.64}$&$ 19.14\pm{1.21}$&$ 0.46\pm{0.01}$&$ 2998\pm {134 }$&$ 4.67\pm{0.09}$&  479/ 461 \\
2200950118 &2019-12-02 & 743  &  58819.41 &  2678 &$ 21.27\pm{1.21}$&$ 15.65\pm{2.01}$&$ 0.45\pm{0.02}$&$ 3094\pm {394 }$&$ 4.51\pm{0.06}$&  402/ 402 \\
2200950119 &2019-12-03 & 2892 &  58820.51 &  2586 &$ 19.99\pm{0.72}$&$ 15.57\pm{1.58}$&$ 0.44\pm{0.01}$&$ 3040\pm {412 }$&$ 4.39\pm{0.07}$&  564/ 504 \\
2200950120 &2019-12-05 & 616  &  58822.19 &  2478 &$ 19.04\pm{0.77}$&$ 13.92\pm{2.06}$&$ 0.44\pm{0.02}$&$ 2747\pm {212 }$&$ 4.51\pm{0.09}$&  352/ 385 \\
2200950121 &2019-12-06 & 1908 &  58823.59 &  2383 &$ 18.78\pm{1.35}$&$ 13.81\pm{2.12}$&$ 0.43\pm{0.02}$&$ 3332\pm {141 }$&$ 4.07\pm{0.04}$&  367/ 354 \\
2200950122 &2019-12-07 & 1795 &  58824.53 &  2311 &$ 18.20\pm{0.86}$&$ 13.41\pm{1.37}$&$ 0.43\pm{0.01}$&$ 3383\pm {258 }$&$ 3.91\pm{0.07}$&  441/ 409 \\
2200950123 &2019-12-08 & 1243 &  58825.80 &  2214 &$ 16.09\pm{0.87}$&$ 13.05\pm{1.58}$&$ 0.42\pm{0.02}$&$ 2926\pm {122 }$&$ 4.50\pm{0.08}$&  407/ 403 \\
2200950124 &2019-12-09 & 2140 &  58826.52 &  2156 &$ 15.78\pm{1.49}$&$ 12.38\pm{1.31}$&$ 0.41\pm{0.02}$&$ 2871\pm {103 }$&$ 4.61\pm{0.09}$&  442/ 432 \\
2200950125 &2019-12-10 & 869  &  58827.56 &  2076 &$ 15.68\pm{2.07}$&$ 12.28\pm{2.11}$&$ 0.41\pm{0.03}$&$ 2893\pm {96  }$&$ 4.05\pm{0.05}$&  413/ 366 \\
2200950126 &2019-12-11 & 1412 &  58828.65 &  2039 &$ 15.62\pm{1.66}$&$ 11.73\pm{1.90}$&$ 0.41\pm{0.02}$&$ 2701\pm {91  }$&$ 5.00\pm{0.03}$&  410/ 368 \\
2200950127 &2019-12-12 & 1247 &  58829.14 &  2000 &$ 14.64\pm{0.79}$&$ 11.71\pm{1.29}$&$ 0.41\pm{0.03}$&$ 3075\pm {136 }$&$ 4.62\pm{0.07}$&  344/ 340  \\
2200950128 &2019-12-13 & 2146 &  58830.33 &  1924 &$ 15.21\pm{0.84}$&$ 11.17\pm{1.38}$&$ 0.41\pm{0.03}$&$ 2901\pm {79  }$&$ 4.30\pm{0.06}$&  315/ 324  \\
2200950129 &2019-12-14 & 3013 &  58831.29 &  1896 &$ 14.16\pm{1.74}$&$ 11.06\pm{1.58}$&$ 0.41\pm{0.02}$&$ 2997\pm {102 }$&$ 5.02\pm{0.04}$&  407/ 383  \\
2200950130 &2019-12-15 & 2077 &  58832.61 &  1853 &$ 12.86\pm{2.91}$&$ 11.61\pm{1.38}$&$ 0.40\pm{0.02}$&$ 2972\pm {145 }$&$ 4.51\pm{0.03}$&  409/ 392  \\
2200950131 &2019-12-16 & 1921 &  58833.45 &  1793 &$ 13.31\pm{1.35}$&$ 11.59\pm{1.47}$&$ 0.40\pm{0.02}$&$ 3210\pm {209 }$&$ 4.51\pm{0.06}$&  410/ 395  \\
2200950132 &2019-12-17 & 1373 &  58834.48 &  1743 &$ 12.66\pm{2.10}$&$ 11.56\pm{1.76}$&$ 0.39\pm{0.02}$&$ 3066\pm {123 }$&$ 4.77\pm{0.05}$&  397/ 356  \\
2200950133 &2019-12-19 & 1338 &  58836.42 &  1673 &$ 11.19\pm{2.40}$&$ 11.19\pm{2.20}$&$ 0.39\pm{0.03}$&$ 2759\pm {140 }$&$ 4.94\pm{0.06}$&  367/ 340  \\
2200950135 &2019-12-22 & 999  &  58839.22 &  1530 &$ 10.64\pm{2.19}$&$ 10.27\pm{2.06}$&$ 0.38\pm{0.03}$&$ 3469\pm {356 }$&$ 4.27\pm{0.08}$&  330/ 314  \\
2200950136 &2019-12-23 & 759  &  58840.06 &  1507 &$ 10.01\pm{1.40}$&$ 10.46\pm{1.77}$&$ 0.38\pm{0.04}$&$ 2872\pm {292 }$&$ 4.73\pm{0.10}$&  365/ 340  \\
2200950137 &2019-12-24 & 854  &  58841.09 &  1462 &$  9.77\pm{0.96}$&$ 10.07\pm{1.68}$&$ 0.38\pm{0.03}$&$ 2843\pm {311 }$&$ 4.97\pm{0.08}$&  381/ 348  \\
2200950138 &2019-12-25 & 954  &  58842.12 &  1428 &$  9.30\pm{0.12}$&$ 10.02\pm{1.51}$&$ 0.37\pm{0.02}$&$ 3118\pm {193 }$&$ 4.65\pm{0.14}$&  371/ 340  \\
2200950139 &2019-12-26 & 963  &  58843.16 &  1398 &$  9.12\pm{0.47}$&$ 9.976\pm{1.994}$&$ 0.38\pm{0.04}$&$ 3202\pm {90 }$&$ 4.55\pm{0.10}$&  317/ 302  \\
2200950140 &2019-12-27 & 903  &  58844.32 &  1356 &$  9.49\pm{0.35}$&$ 9.525\pm{0.649}$&$ 0.38\pm{0.01}$&$ 2883\pm {213}$&$ 4.35\pm{0.07}$&  471/ 447  \\
2200950142 &2019-12-29 & 1076 &  58846.66 &  1275 &$  7.95\pm{0.60}$&$ 9.976\pm{1.322}$&$ 0.37\pm{0.03}$&$ 3248\pm {141}$&$ 4.49\pm{0.05}$&  385/ 332  \\
2200950144 &2020-01-01 & 1726 &  58849.23 &  1197 &$  7.35\pm{0.74}$&$ 9.715\pm{1.108}$&$ 0.36\pm{0.01}$&$ 3375\pm {293}$&$ 4.62\pm{0.11}$&  469/ 372  \\
2200950145 &2020-01-02 & 1873 &  58850.26 &  1171 &$  6.93\pm{1.73}$&$ 9.599\pm{1.034}$&$ 0.35\pm{0.02}$&$ 2998\pm {144}$&$ 4.64\pm{0.11}$&  402/ 356  \\
2200950146 &2020-01-03 & 1092 &  58851.23 &  1143 &$  7.24\pm{0.58}$&$ 8.778\pm{1.506}$&$ 0.35\pm{0.03}$&$ 3440\pm {196}$&$ 4.80\pm{0.14}$&  348/ 318  \\
2200950147 &2020-01-04 & 1333 &  58852.46 &  1106 &$  6.77\pm{1.24}$&$ 8.075\pm{1.244}$&$ 0.35\pm{0.04}$&$ 2983\pm {144}$&$ 4.93\pm{0.12}$&  370/ 325  \\
2200950148 &2020-01-05 & 1166 &  58853.49 &  1093 &$  6.91\pm{4.21}$&$ 7.040\pm{1.417}$&$ 0.34\pm{0.02}$&$ 3361\pm {121}$&$ 4.50\pm{0.18}$&  326/ 320  \\
2200950149 &2020-01-06 & 1091 &  58854.65 &  1069 &$  6.44\pm{1.36}$&$ 6.975\pm{1.145}$&$ 0.33\pm{0.03}$&$ 2910\pm {74 }$&$ 4.65\pm{0.19}$&  366/ 326  \\
2200950151 &2020-01-08 & 2725 &  58856.59 &  1020 &$  6.27\pm{0.95}$&$ 7.374\pm{0.733}$&$ 0.32\pm{0.02}$&$ 3136\pm {92 }$&$ 4.63\pm{0.10}$&  372/ 387  \\
2200950152 &2020-01-09 & 1493 &  58857.40 &  1008 &$  6.76\pm{3.12}$&$ 7.094\pm{1.293}$&$ 0.31\pm{0.03}$&$ 2925\pm {110}$&$ 4.33\pm{0.06}$&  391/ 372  \\
2200950153 &2020-01-10 & 991  &  58858.37 &   984 &$  5.82\pm{2.38}$&$ 6.651\pm{0.483}$&$ 0.30\pm{0.02}$&$ 3276\pm {169}$&$ 4.16\pm{0.10}$&  381/ 356  \\
2200950154 &2020-01-14 & 503  &  58862.85 &   133 &$  0.66\pm{0.13}$&$ 0.822\pm{0.131}$&$ 0.17\pm{0.02}$&$ 10446\pm{799}$&$ 2.38\pm{0.12}$&  192/ 187  \\
2200950155 &2020-01-21 & 1107 &  58869.74 &   142 &$  0.26\pm{0.82}$&$ 1.173\pm{0.094}$&$ 0.13\pm{0.03}$&$ 10639\pm{1494}$&$ 1.91\pm{0.07}$&  317/ 331 \\
2200950156 &2020-01-23 & 2283 &  58871.58 &   104 &$  0.15\pm{0.03}$&$ 0.991\pm{0.122}$&$ 0.12\pm{0.02}$&$ 15083\pm{2212}$&$ 1.89\pm{0.05}$&  364/ 401 \\
2200950157 &2020-01-24 & 2419 &  58872.12 &  86.4 &$  0.12\pm{0.01}$&$ 0.937\pm{0.093}$&$ 0.13\pm{0.01}$&$  9525\pm{1283}$&$ 1.88\pm{0.04}$&  400/ 412 \\
2200950158 &2020-01-25 & 908  &  58873.19 &  70.4 &$  0.07\pm{0.01}$&$ 0.718\pm{0.085}$&$ 0.12\pm{0.02}$&$ 10751\pm{3212}$&$ 1.81\pm{0.06}$&  239/ 258 \\
2200950159 &2020-01-26 & 2511 &  58874.61 &  61.8 &$  0.05\pm{0.02}$&$ 0.475\pm{0.121}$&$ 0.11\pm{0.02}$&$ 13211\pm{2539}$&$ 1.85\pm{0.09}$&  410/ 394 \\
2200950160 &2020-01-28 & 1258 &  58876.36 &  47.9 &$  0.08\pm{0.01}$&$ 0.686\pm{0.112}$&$ 0.13\pm{0.01}$&$  8477\pm{1677}$&$ 1.83\pm{0.05}$&  324/ 281 \\
2200950161 &2020-01-29 & 1444 &  58877.58 &  29.7 &$  0.03\pm{0.01}$&$ 0.514\pm{0.078}$&$ 0.11\pm{0.02}$&$ 14928\pm{2823}$&$ 1.75\pm{0.07}$&  316/ 267 \\
2200950162 &2020-01-30 & 1356 &  58878.42 &  27.7 &$     -         $&$ 0.515\pm{0.049}$&$    -          $&$          -   $&$ 1.89\pm{0.13}$&  291/ 261 \\
\hline
\end{tabular}
\end{table*}

\begin{table*}
\contcaption{Best-fit parameters obtained from spectral fitting of data obtained from the {\it NICER} observations of MAXI~J0637--430 during its 2019-2020 outburst}
\begin{tabular}{lccccccccccccc}
\hline
Obs ID & UT Date & Exp & Date & Count Rate & DBB & PL Flux$^a$ & $T_{\rm in}$ & $N_{\rm DBB}$ & $\Gamma$ &$\chi^2$/dof \\
 &(yyyy-mm-dd) &(s) &(MJD) & (Count s$^{-1}$) &Flux$^a$  &  &   (keV)& & &  \\
\hline
2200950163 &2020-01-31 & 2874 &  58879.55 &  23.4 &$     -         $&$ 0.512\pm{0.043}$&$    -          $&$          -    $&$ 1.88\pm{0.09}$&  367/356  \\
2200950164 &2020-02-01 & 6938 &  58880.59 &  21.6 &$     -         $&$ 0.282\pm{0.081}$&$    -          $&$          -    $&$ 1.84\pm{0.10}$&  541/ 484 \\
2200950165 &2020-02-02 & 4619 &  58881.39 &  18.4 &$     -         $&$ 0.338\pm{0.042}$&$    -          $&$          -    $&$ 1.88\pm{0.15}$&  383/ 351 \\
2200950166 &2020-02-03 & 1563 &  58882.43 &  16.2 &$     -         $&$ 0.279\pm{0.038}$&$    -          $&$          -    $&$ 1.86\pm{0.10}$&  283/ 245 \\
2200950167 &2020-02-04 & 1728 &  58883.72 &  13.9 &$     -         $&$ 0.287\pm{0.027}$&$    -          $&$          -    $&$ 1.78\pm{0.09}$&  189/220  \\
2200950168 &2020-02-05 & 683  &  58884.11 &  11.0 &$     -         $&$ 0.022\pm{0.002}$&$    -          $&$          -    $&$ 1.84\pm{0.10}$&  541/498  \\
2200950169 &2020-02-06 & 1314 &  58885.33 &  13.9 &$     -         $&$ 0.031\pm{0.003}$&$    -          $&$          -    $&$ 1.65\pm{0.12}$&  271/264 \\
2200950170 &2020-02-07 & 629  &  58886.63 &   9.1 &$     -         $&$ 0.015\pm{0.004}$&$    -          $&$          -    $&$ 1.94\pm{0.08}$&  102/96  \\
2200950171 &2020-02-08 & 912  &  58887.40 &  11.7 &$     -         $&$ 0.021\pm{0.002}$&$    -          $&$          -    $&$ 2.07\pm{0.04}$&  197/173 \\
2200950173 &2020-02-10 & 1285 &  58889.61 &  11.5 &$     -         $&$ 0.023\pm{0.004}$&$    -          $&$          -    $&$ 1.84\pm{0.23}$&  216/183 \\
2200950174 &2020-02-11 & 884  &  58890.81 &  10.8 &$     -         $&$ 0.021\pm{0.003}$&$    -          $&$          -    $&$ 1.86\pm{0.27}$&  144/137 \\
2200950175 &2020-02-13 & 934  &  58892.10 &   9.4 &$     -         $&$ 0.018\pm{0.003}$&$    -          $&$          -    $&$ 1.92\pm{0.18}$&  127/136 \\
2200950176 &2020-02-14 & 1550 &  58893.55 &   9.1 &$     -         $&$ 0.018\pm{0.002}$&$    -          $&$          -    $&$ 1.89\pm{0.14}$&  155/179 \\
2200950177 &2020-02-16 & 1392 &  58895.68 &   8.2 &$     -         $&$ 0.016\pm{0.002}$&$    -          $&$          -    $&$ 1.97\pm{0.18}$&  179/158 \\
2200950178 &2020-02-17 & 233  &  58896.31 &   8.0 &$     -         $&$ 0.016\pm{0.003}$&$    -          $&$          -     $&$ 1.82\pm{0.19}$&  34/39  \\
2200950181 &2020-02-20 & 777  &  58899.73 &   6.2 &$     -         $&$ 0.011\pm{0.003}$&$    -          $&$          -     $&$ 2.22\pm{0.19}$&  61/50  \\
2200950182 &2020-02-21 & 1129 &  58900.39 &   7.2 &$     -         $&$ 0.016\pm{0.002}$&$    -          $&$          -     $&$ 1.68\pm{0.21}$&  111/105\\
2200950183 &2020-02-22 & 1391 &  58901.04 &   7.0 &$     -         $&$ 0.014\pm{0.002}$&$    -          $&$          -     $&$ 1.84\pm{0.14}$&  131/160\\
2200950184 &2020-02-23 & 1538 &  58902.17 &   6.3 &$     -         $&$ 0.012\pm{0.003}$&$    -          $&$          -     $&$ 1.95\pm{0.14}$&  125/141\\
2200950185 &2020-02-24 & 1035 &  58903.59 &   5.0 &$     -         $&$ 0.009\pm{0.003}$&$    -          $&$          -     $&$ 1.88\pm{0.09}$&  239/224 \\
2200950186 &2020-02-25 & 842  &  58904.31 &   5.2 &$     -         $&$ 0.009\pm{0.002}$&$    -          $&$          -     $&$ 2.46\pm{0.12}$&  80/70  \\
2200950187 &2020-02-27 & 1352 &  58906.07 &   5.1 &$     -         $&$ 0.009\pm{0.002}$&$    -          $&$          -     $&$ 2.16\pm{0.15}$&  135/97 \\
2200950188 &2020-02-28 & 1934 &  58907.21 &   5.2 &$     -         $&$ 0.009\pm{0.001}$&$    -          $&$          -     $&$ 1.94\pm{0.13}$&  127/128 \\
2200950189 &2020-02-29 & 1588 &  58908.11 &   5.4 &$     -         $&$ 0.010\pm{0.002}$&$    -          $&$          -     $&$ 1.87\pm{0.13}$&  56/53  \\
3200950101 &2020-03-01 & 1230 &  58909.15 &   4.2 &$     -         $&$ 0.007\pm{0.001}$&$    -          $&$          -     $&$ 2.24\pm{0.16}$&  134/111 \\
3200950102 &2020-03-02 & 1610 &  58910.35 &   4.7 &$     -         $&$ 0.008\pm{0.002}$&$    -          $&$          -     $&$ 2.02\pm{0.07}$&  138/121 \\
3200950103 &2020-03-03 & 1085 &  58911.57 &   4.4 &$     -         $&$ 0.008\pm{0.002}$&$    -          $&$          -     $&$ 2.12\pm{0.11}$&  92/91 \\
3200950104 &2020-03-04 & 231  &  58912.51 &   4.1 &$     -         $&$ 0.008\pm{0.003}$&$    -          $&$          -     $&$ 2.04\pm{0.10}$&  194/166 \\
3200950105 &2020-03-05 & 287  &  58913.09 &   3.9 &$     -         $&$ 0.007\pm{0.002}$&$    -          $&$          -     $&$ 2.06\pm{0.09}$&  28/33   \\
3200950108 &2020-03-16 & 997  &  58924.21 &   2.9 &$     -         $&$ 0.007\pm{0.002}$&$    -          $&$          -     $&$ 1.86\pm{0.11}$&  27/32  \\
3200950109 &2020-03-22 & 947  &  58930.51 &   2.4 &$     -         $&$ 0.004\pm{0.001}$&$    -          $&$          -     $&$ 1.91\pm{0.10}$&  51/57  \\
3200950110 &2020-03-24 & 1370 &  58933.02 &   2.3 &$     -         $&$ 0.004\pm{0.002}$&$    -          $&$          -     $&$ 2.21\pm{0.06}$&  66/55 \\
3200950111 &2020-04-08 & 1240 &  58947.09 &   1.5 &$     -         $&$ 0.002\pm{0.001}$&$    -          $&$          -     $&$ 2.62\pm{0.09}$&  55/45 \\

\hline
\end{tabular}
\leftline{$^a$ Fluxes are in the unit of 10$^{-10}$ erg cm$^{-2}$ s$^{-1}$.}
\leftline{Errors are quoted at 90\% confidence.}
\end{table*}



\bsp	
\label{lastpage}
\end{document}